\documentclass[twocolumn,letterpaper,aps,prd,superscriptaddress,longbibliography,showpacs,floatfix,preprintnumbers]{revtex4-2}
\usepackage{color}
\usepackage{graphicx}	
\usepackage{xspace}     
\usepackage{hyperref}
 \usepackage{orcidlink}

\newcommand{\xp}{\mbox{$x_p$}\xspace}
\begin{document}


\title{Production cross sections of light and charmed mesons \\in $e^+e^-$ annihilation near 10.58 GeV }

\date{\today}

\begin{abstract}
We report measurements of production cross sections for $\rho^+$, $\rho^0$, $\omega$, $K^{*+}$, $K^{*0}$, $\phi$, $\eta$, $K_S^0$, $f_0(980)$, $D^+$, $D^0$, $D_s^+$, $D^{*+}$, $D^{*0}$, and $D^{*+}_s$ 
in $e^+e^-$ collisions at a center-of-mass energy near 10.58 GeV. The data were recorded by the Belle experiment, consisting of 571 fb$^{-1}$ at 10.58 GeV and 74 fb$^{-1}$ at 10.52 GeV. Production cross sections are extracted as a function of the fractional hadron momentum \xp . 
The measurements are compared to {\sc pythia} Monte Carlo generator predictions with various fragmentation settings, including those that have increased fragmentation into vector mesons over pseudo-scalar mesons. The cross sections measured for light hadrons are consistent with no additional increase of vector over pseudo-scalar mesons. 
The charmed-meson cross sections are compared to earlier measurements---when available---including older Belle results, which they supersede. They are in agreement before application of an improved initial-state radiation correction procedure that causes slight changes in their \xp shapes.    

\end{abstract}

\noaffiliation
 \author{R.~Seidl\,\orcidlink{0000-0002-6552-6973}} 
 
  \author{I.~Adachi\,\orcidlink{0000-0003-2287-0173}} 
  \author{H.~Aihara\,\orcidlink{0000-0002-1907-5964}} 
  \author{T.~Aushev\,\orcidlink{0000-0002-6347-7055}} 
  \author{R.~Ayad\,\orcidlink{0000-0003-3466-9290}} 
  \author{Sw.~Banerjee\,\orcidlink{0000-0001-8852-2409}} 
  \author{K.~Belous\,\orcidlink{0000-0003-0014-2589}} 
  \author{J.~Bennett\,\orcidlink{0000-0002-5440-2668}} 
  \author{M.~Bessner\,\orcidlink{0000-0003-1776-0439}} 
  \author{B.~Bhuyan\,\orcidlink{0000-0001-6254-3594}} 
% ...
\collaboration{The Belle Collaboration}

\maketitle

\section{Introduction}

Fragmentation functions (FFs) describe the formation of hadrons, states of confined quarks and gluons (collectively denoted partons), out of asymptotically free, highly energetic partons. As they cannot be calculated from first principles in quantum chromodynamics they need to be extracted experimentally. These nonperturbative objects in turn can be used to describe hadron-production cross sections in various high-energy processes since the nonperturbative fragmentation functions (and potentially other nonperturbative objects, like parton distribution functions) factorize from the calculable hard interactions \cite{Collins:2011zzd}. High-precision knowledge of FFs can thus provide additional sensitivity to the flavor, spin, and transverse momentum of partons, e.g., in measurements of parton distributions and therefore to the three-dimensional and spin-structure of the nucleon, one of the key scientific goals of the future electron-ion collider (EIC) \cite{Accardi:2012qut,EICYellowReport}. 
Given that in electron-positron annihilation no hadrons exist in the initial state, such collisions are well suited for obtaining fragmentation functions as they are the only nonperturbative functions that need to be modeled. 

So far, many measurements have been performed at various collision energies for light and charmed hadron production \cite{TASSO:1980dyh,TASSO:1982bkc,TPCTwoGamma:1988yjh,ALEPH:1994cbg,DELPHI:1998cgx,SLD:2003ogn,BaBar:2013yrg,Belle:2013lfg,Belle:2005mtx,CLEO:2004enr,BaBar:2002ncl,BESIII:2022zit,BESIII:2024hcs}. What has not been studied intensively to date are fragmentation functions for the various vector mesons (VMs) \cite{Abreu:1998nn,Ackerstaff:1998ue}, even though they have recently gained increased interest. Given their slightly higher masses compared to their pseudoscalar counterparts, the production cross sections might be reduced at low energies, but in high-energy processes no a priori reason exists that suggests they would be suppressed. Many Monte Carlo (MC) generators such as {\sc pythia} \cite{Sjostrand:2001yu,Sjostrand:2000wi} treat the fragmentation of light partons into pseudoscalar and vector mesons equally apart from their mass dependence. 

Understanding the production of VMs can address
various open questions. For example, the muon puzzle \cite{Albrecht:2021cxw}
in modeling ultra-high-energetic cosmic-ray air-shower
measurements can be related to the relative strength
of the production of neutral rho mesons vs.\ neutral pions
as it affects the hadronic-shower evolution \cite{Ostapchenko:2013pia,Riehn:2019jet}, and
thus the muon content of the cosmic-ray air shower. 
Similarly, while strong decays are explicitly included in the formal description of FFs, the decay contributions to pion and kaon FFs may exhibit significantly different transverse-momentum dependencies as compared, e.g., to the contributions from directly produced pions and kaons, particularly at low momenta. 
Last, the transverse-spin dependent Collins FFs \cite{Collins:1992kk} are expected to behave differently for VMs than for pseudoscalar mesons in various models \cite{Artru:2012zz,Artru:1974hr,Kerbizi:2021gos}, due to the way the polarized quark has to combine in the fragmentation process to create either a spin-zero or spin-one meson. While few measurements related to VM Collins asymmetries exist \cite{Alexeev:2022wgr}, the measurements presented here will provide the unpolarized baseline for such studies.  \\

Additionally, we present an update of \(D\)-meson cross sections as a function of the fractional hadron momentum that confirm and supersede previous Belle measurements \cite{Belle:2005mtx} and that are also compared to other published results. Furthermore, $D_s^{*+}$ meson cross sections are presented for the first time. The formation of heavy-flavor mesons is of general interest at high-energy colliders and particularly in heavy-ion collisions, where the formation of heavy-flavor mesons may be modified by the nuclear environment or the quark-gluon plasma. At the EIC, \(D\) mesons in the final state will be prime candidates to single out gluonic hard interactions and thus help to access the gluon (spin and momentum) structure of the nucleon \cite{EICYellowReport}. 

Last, also the cross sections for several other light pseudoscalar mesons are reported since those are produced in abundance in high-energy processes and provide complementary sensitivity, particularly to strange quarks even in experimental setups where charged kaons cannot be directly identified. 

The measurements reported here will help improve the precision of FF extractions that are needed as input in all of these mentioned processes.   
Since the high precision of recent Belle fragmentation-related measurements \cite{Belle:2020pvy} has challenged the flexibility of various global fragmentation fits, we provide (online) all sources of uncertainties separately, together with the information on whether they are global scale uncertainties, correlated among bins or uncorrelated.  

This paper is organized as follows. After this introduction, in section \ref{sec:data}, the data sets and MC simulations are discussed. Section \ref{sec:event} details the event and particle selection. In section \ref{sec:corr}, the invariant-mass fits and the subsequent acceptance and efficiency corrections are described. Section \ref{sec:systematics} covers all systematic consistency tests and summarizes the total systematic uncertainty budgets of these measurements, before the final results are presented in section \ref{sec:results} and summarized in section \ref{sec:summary}. 
\section{Data sets and Monte Carlo simulations \label{sec:data}}
The data were taken with the Belle detector \cite{belle,Brodzicka:2012jm} at the asymmetric electron-positron collider KEKB \cite{KEKB,Abe:2013kxa}.
 Here, 8~GeV electrons collided with 3.5~GeV positrons at the center-of-mass energy of the $\Upsilon$(4S) resonance at 10.58 GeV (denoted on-resonance throughout this manuscript).  We also take data at 60 MeV below the $\Upsilon(4S)$ resonance (denoted as continuum). In addition to the open quark-antiquark pair production (at leading order) and other QED processes, the on-resonance data also contain events from $\Upsilon(4S)$ production with their subsequent decays into $B$-meson pairs. The accumulated data sets consist of 558 fb$^{-1}$ in the on-resonance sample and 74 fb$^{-1}$ in the continuum sample. The two samples are initially analyzed separately in order to compare the results for consistency and to see the contribution of \(B\) decays to the cross sections.

The events were collected with the Belle detector, which is a large-solid-angle magnetic spectrometer comprising of a silicon vertex detector (SVD),
a 50-layer central drift chamber, an array of aerogel threshold Cherenkov counters (ACC), a barrel-like arrangement of time-of-flight (TOF) scintillation counters, and an electromagnetic calorimeter (ECL) made of CsI(Tl) crystals located inside a superconducting solenoid coil that provides a
1.5 T magnetic field. An iron flux-return located outside
of the coil is instrumented to detect $K^0_{L}$ mesons and to
identify muons (KLM). The detector is described in detail in publications \cite{belle} and \cite{Brodzicka:2012jm}. The data for this measurement were collected in a detector configuration consisting of a beampipe with a 1.5 cm radius and 1 mm thickness, a 4-layer SVD, and a small-cell inner drift chamber.

For simulations, {\sc pythia}6 \cite{Sjostrand:2006za} was used as a generator for the $e^+e^- \rightarrow q \bar{q}$ events, where $q$ is a $u$, $d$, $s$, or $c$ quark (denoted as udsc MC and specified as the \textit{old Belle tune} in the Data-MC comparisons), as well as other QED processes. This generator is included in the package {\sc evtgen} \cite{evtgen}, which also handles the $\Upsilon(4S)$ production and the subsequent decays. Additionally, dedicated $\tau$-pair simulations were obtained using the generator {\sc kkmc} \cite{aafh} and the {\sc tauola} package \cite{tauola}. All generated events were fed into full {\sc geant}3 \cite{Brun:1994aa} detector simulations and the event reconstruction for further analysis. 
For the quark-pair production events, which are the main source for the FF information this analysis provides, additional events were generated with various fragmentation parameter settings in {\sc pythia}, hereafter denoted as tunes (see Table \ref{tab:jetsettable}).
The latter events provide a tool to evaluate the systematic effects of these tunes on acceptance and initial-state radiation corrections, as well as provide information regarding the most suitable fragmentation settings for the generator. All simulations are performed at \(\sqrt{s}=10.58\) GeV as the difference in center-of-mass energies has a negligible effect on all correction steps.

\section{Event and particle selection criteria\label{sec:event}}
Hadronic events are selected with a visible-energy requirement of at least 7 GeV, obtained from the sum of the momenta of reconstructed tracks and energies of trackless calorimeter clusters, in order to reduce the contamination from $e^+e^-\rightarrow\tau^+\tau^-$ events. The heavy-jet mass is defined as the greater of the invariant mass sums of all particles in one hemisphere as generated by the plane perpendicular to the thrust axis \cite{Brandt:1964sa}. It is required that the heavy-jet mass lies above 1.8 GeV or that the ratio of the heavy-jet mass to the visible energy lies above 0.25. These criteria reduce the amount of two-photon processes. Additionally, at least three charged tracks have to be reconstructed, which also reduces the contamination of two-photon processes, as well as lepton-pair production. 

A charged particle is selected if the track emerges from a region that is less than 4 cm away from the nominal interaction point along the positron beam direction and less than 2 cm away perpendicularly (except for $K_S^0$ candidates).  
Tracks considered also need to be reconstructed in the barrel region of the detector where $-0.511 < \cos \theta_{{\text{Lab}}} < 0.842 $ and information on particle identification is available from all relevant detectors that contribute to its determination. A minimum transverse momentum of 50 MeV \footnote{Throughout this manuscript $c$ is set to unity.} is required to ensure the tracks traverse all particle identification (PID) detectors. 
The information from the ACC, TOF, drift chamber, ECL, and KLM is then used to identify the particles using likelihood ratios between pion, kaon, and proton hypotheses, as well as muon or electron likelihoods. Particle misidentification is corrected as described in Ref.~\cite{Belle:2013lfg} using a predominantly data-based calibration of the above five particle types in a fine 17 $\times$ 9 laboratory-momentum and polar-angular binning. For momenta below 500 MeV, the particle type given by the PID detectors is assumed to be correct as the flight times are sufficiently long. In regions where the phase space requires particles with these lower laboratory momenta (to be discussed in further detail below), such particles are also considered but require an additional reconstruction efficiency correction. This predominantly affects the slow charged pion of the $D^{*+}\rightarrow D^0\pi^+$ decay, as well as the region of low fractional momentum for a few of the mesons considered here. 

Neutral-pion candidates are selected by the electromagnetic calorimeters if the invariant mass of a pair of trackless clusters falls between 120 and 150 MeV, which is more than approximately two standard deviations on the di-photon mass peak at Belle. Also a minimum transverse momentum of 50 MeV is required for each of the candidate photons in order to reduce the combinatorial background. Despite the larger acceptance of the calorimeter, for most reconstructed pions the same acceptance and momentum requirements as for charged hadrons are applied for consistency reasons, with the same exception for the slow pion of the $D^{*0}\rightarrow D^0 \pi^0$ decay and other final states where the phase space requires slow pions.

Once the produced particles are identified, they are organized into either pairs or triplets for the decay channels $\rho^0\rightarrow \pi^+\pi^-$, $\rho^\pm\rightarrow \pi^\pm\pi^0$, $\omega\rightarrow \pi^+\pi^-\pi^0$, $K^{*0}\rightarrow K^+\pi^-$, $K^{*\pm}\rightarrow K^\pm\pi^0$, $\phi\rightarrow K^+K^-$, as well as for the $K_S^0 \rightarrow \pi^+\pi^-$, $\eta \rightarrow \pi^+\pi^-\pi^0$, and $f_0(980) \rightarrow \pi^+\pi^-$ channels. Apart from the $\rho$ mesons, charge-conjugate states are combined and are implicitly analyzed together.  
For $K_S^0$ decay candidates, the transverse decay vertex position as obtained from an additional vertex fit within 30 MeV of the nominal $K_S^0$ mass has to be less than 5 cm away to account for the finite lifetime of this particle.
For the \(D\) mesons, the following decay channels are analyzed: $D^{0}\rightarrow K^-\pi^+$, $D^{+}\rightarrow K^-\pi^+\pi^+$, $D^{*+}\rightarrow D^0 \pi^+ \rightarrow (K^-\pi^+)\pi^+$, 
$D^{*0}\rightarrow D^0 \pi^0 \rightarrow (K^-\pi^+)\pi^0$, $D_s^{+}\rightarrow \phi \pi^+ \rightarrow (K^+K^-)\pi^+$, and $D_s^{*+}\rightarrow \phi \pi^+ \gamma \rightarrow (K^+K^-)\pi^+ \gamma$. Additional $D^0$ and $D_s$ final states are also considered, but only used for consistency tests. For the $D_s^*$ final state, the decay photon is required to be recorded as a trackless cluster with a minimum transverse momentum of 50 MeV.  

The various final states are binned in 40 equidistant bins of $x_p = p_h /p_{{\mathrm{ max}}}$, which is the fractional hadron momentum that takes into account the maximal momentum available for a particle of a certain mass, $p_{\mathrm{ max}} = \sqrt{s/4  - m_h^2}$, where $\sqrt{s}$ is the total center-of-mass energy and $m_h$ is the nominal hadron mass taken from the particle data group (PDG) \cite{ParticleDataGroup:2024cfk}. Unlike the hadron fractional energy \(z\) (defined with respect to half the center-of-mass energy) that is often used for fragmentation functions, $x_p$ is a true scaling variable that runs from zero to unity. 
For very low masses or very high center-of-mass energies, the two definitions are nearly identical, but for \(B\) factory energies they are substantially different. The PDG masses have been used to calculate $p_{\mathrm {max}}$ (and therefore \xp) also in the case of the wider resonances for which the invariant mass of the decay particles can significantly differ from the nominal mass of the parent. It is found that the differences in \xp even for the wide resonances are insignificant relative to the bin widths.

\begin{figure*}[htb]
    \centering
    \includegraphics[width=0.98\textwidth]{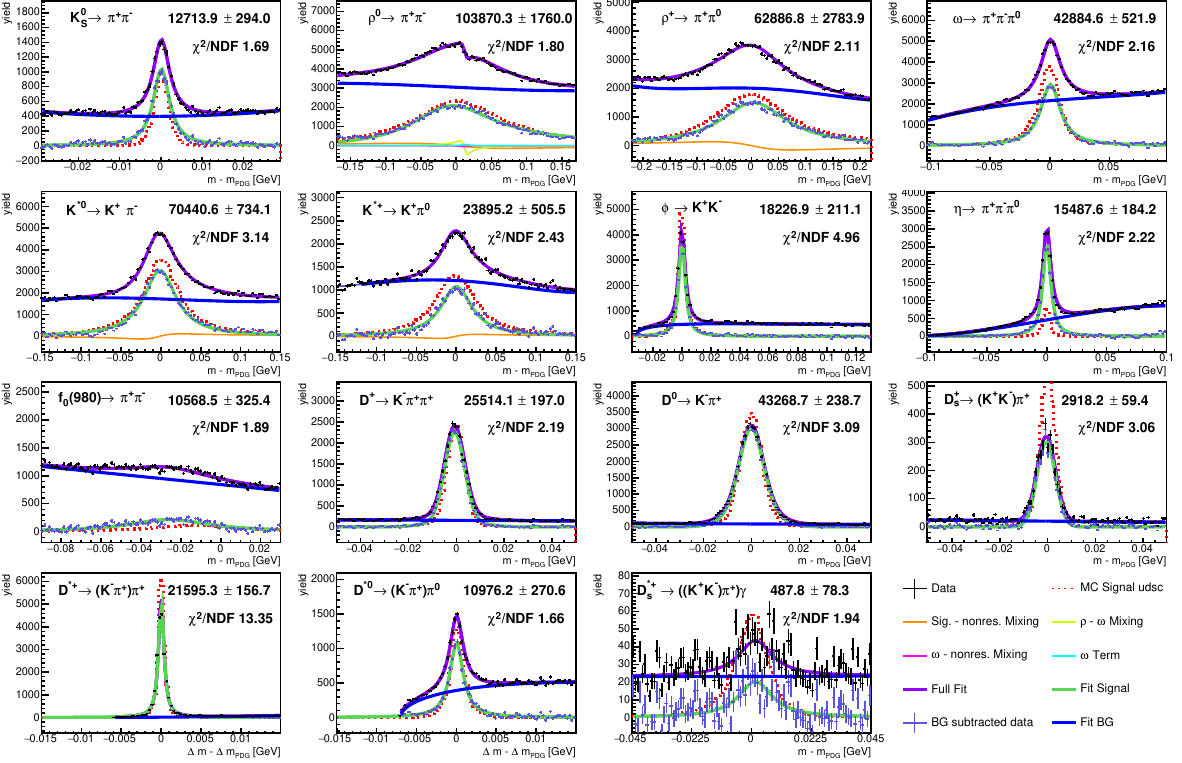}
    \caption{Examples of the invariant-mass fits for an intermediate \xp bin (0.625--0.650) for all decay channels considered. In addition to the continuum data (black points), also the background subtracted data (blue points) and MC signal histograms (red, dashed lines) 
    are displayed. The total fit results are shown as violet lines, while the signal parts are displayed in green and the background parts in blue. Where interference terms are considered, they are also displayed (signal to non-resonant background in orange, $\omega$ to non-resonance background in magenta, $\rho$-$\omega$ mixing in light green, and $\omega$ signal in cyan lines). Each panel also displays the signal yield \(N_{\text{Fit}}\) as well as the fit quality over the whole mass range used.}
    \label{fig:massfits}
\end{figure*}

\section{Signal yields and corrections\label{sec:corr}}

\subsection{Invariant-mass fits}
In each \xp bin, the invariant mass of the various mesons under study is calculated in 100 mass bins that cover between $\pm$50 MeV and $\pm$220 MeV around the nominal PDG values, depending on the width of the resonance. For the two $D^*$ states, the mass differences between $D^*$ and $D^0$ are binned instead because of the small momentum of the slow pion. For intermediate particles such as the $D^0\rightarrow K \pi $ from the $D^*$ decays or the $\phi \rightarrow K K $ from the $D_s$ decays, an additional mass window of 15~MeV and 7~MeV was applied, respectively, which corresponds to between one and two standard deviations of the mass peaks. 

To obtain the signal yields, varying signal and background functions are fitted to the invariant-mass distributions for each final state and in each \xp bin. The functional forms for the background are chosen for most resonances to be first- to third-order polynomials that best describe the background in the MC in terms of $\chi^2$ per degree of freedom. For the $D^*$ and $\phi$ backgrounds, a threshold function of the form $\left[A + B(m  - m_0)\right]^{[C+D(m  - m_0)]}$ is used, where $m_0$ is the threshold value, and \(A\), \(B\), \(C\), as well as \(D\) are free parameters.

The signal functions are either relativistic Breit-Wigner or Gaussian shapes, where again the functional form is determined by the MC signal distributions. Generally, magnitude, mass, and width are not constrained.
In the actual fits to the data, the best parameters of the MC-based signal and background fits are used as initial parameters but are generally allowed to vary freely, since the absolute level of background is not perfectly described  by the MC simulations in all \xp bins, as already observed in, e.g., Ref.~\cite{Seidl:2017qhp}. The central mass of the signal and its width can be constrained based on MC, but that does not generally impact the result of the fits. Such constraints can be modestly helpful in the very low \xp region for fits with large background. 

In the case of neutral $\rho$ mesons, the interference with the $\omega \rightarrow \pi\pi$ decay as well as with the non-resonant two-pion background needs to be taken into account \cite{Soeding:1966zz,STAR:2007elq}. 
Here, the fit function is 
\begin{widetext}
\begin{equation}    
    \frac{dN^{\pi\pi}}{dm_{\pi\pi}} = \left| A \frac{\sqrt{m_{\pi\pi}m_\rho \Gamma_\rho}}{m^2_{\pi\pi} - m^2_\rho + i m_\rho \Gamma_\rho} + B + C e^{i\phi}         \frac{\sqrt{m_{\pi\pi}m_\omega \Gamma_\omega}}{m^2_{\pi\pi} - m^2_\omega + i m_\omega \Gamma_\omega} \right|^2
    \label{eq:interf}
\end{equation}
\end{widetext}
with Breit-Wigner forms for the two resonances and a constant for the non-resonant part. The widths $\Gamma_{\rho,\omega}$ are defined by 
\begin{eqnarray}
\Gamma_\rho = \Gamma^0_\rho \frac{m_\rho}{m_{\pi\pi}} \left( \frac{m^2_{\pi\pi} - 4 m^2_\pi}{m^2_\rho - 4 m^2_\pi}\right)^{3/2} \\
\Gamma_\omega = \Gamma^0_\omega \frac{m_\omega}{m_{\pi\pi}} \left( \frac{m^2_{\pi\pi} - 4 m^2_\pi}{m^2_\omega - 4 m^2_\pi}\right)^{3/2}.
\label{eq:interf2}
\end{eqnarray}
All three amplitudes $A$, $B$, and $C$, the $\rho - \omega$ interference phase $\phi$, as well as the masses and widths of the \(\rho\) and \(\omega\) are used as parameters, in addition to the background. Similarly, for charged $\rho$ and $K^*$, a potential interference with a non-resonant background was included in the fit functions. Its impact on the signal is marginal but it does improve the overall quality of the fit. 

The fit procedure outlined above can generally describe the data very well, as shown in Fig.~\ref{fig:massfits} for an example \xp bin and all final-state particles considered here. In general, the level of background decreases with increasing \xp and is also smaller the heavier the meson considered. For all particles that contain a neutral pion as a decay product, the background level is generally higher due to the additional background in the $\pi^0$ reconstruction coming from pairs of uncorrelated photons. 
We also note that because $D$ meson FFs peak at large \xp, their signal contributions can be relatively small at low $x_p$, leading to
large uncertainties in the extracted values.  

To estimate the reliability of the signal yields thus obtained, several ways to extract the signal are considered. In addition to calculating the yield from the integral over the signal part of the fit function, also the integral of the histogram after subtracting the fitted background parameterization is calculated. The background functional form is varied to gauge the sensitivity to the background by using those fits that describe the MC background second and third-best as indicated by the reduced $\chi^2$. These variations are all considered as sources of systematic uncertainty related to the signal extraction.

In the next step, the extracted yields \(N_{\text{Fit}}\) are corrected for the branching fractions based on the PDG \cite{ParticleDataGroup:2024cfk} values, tabulated in Table \ref{tab:BRs} for reference. In the case of the $K^*$ decays, the fraction decaying into $K^0 \pi$ rather than the charged-kaon related final states studied here are corrected in the following acceptance-correction step because the PDG does not separate the two decay modes (hence our correction uses the MC-based branching fractions). Similarly, contributions from tails of the very wide resonances that exceed the mass windows used for the fits, most notably the $\rho$ mesons, are also dealt with during the acceptance correction step.    
\begin{figure*}[htb]
    \centering
    \includegraphics[width=0.9\textwidth]{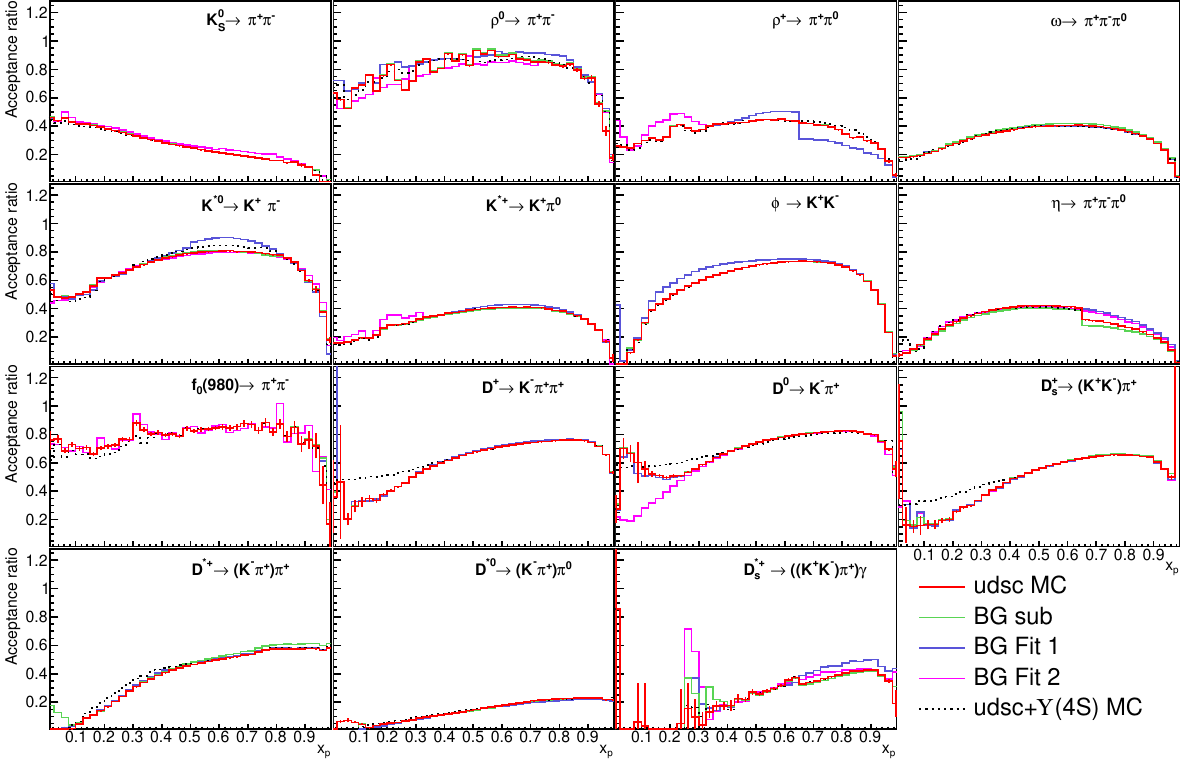}
    \caption{Acceptance and reconstruction efficiencies within the barrel part of the detector as a function of \xp for the various particles considered. The differently colored points show results from the various methods of extracting the signal yields (labeled BG sub for the method of just subtracting the background function, and BG Fit 1/2, for the two additional functional forms, considered) from the invariant-mass distributions, based on the quality of the MC fits for different functional forms. Those are used correspondingly in correcting their respective yields. The dashed black lines correspond to the efficiencies obtained in MC data samples including $\Upsilon (4S)$ decays.}
    \label{fig:acc1}
\end{figure*}

Before proceeding further to the acceptance corrections that follow, the non-$q\bar{q}$ contributions, estimated from MC calculation, are subtracted from the signal yields in the final step of this stage. These contributions were obtained using the aforementioned two-photon and $e^+e^-\rightarrow\tau^+\tau^-$ MC events analyzed in the same way as the data but using the true signal contributions instead of fits, except for final states including neutral pions where the true pion association was not always available. 

\subsection{Acceptance and efficiency correction}
The \xp binning chosen for this analysis is much coarser than the \xp resolution (below $0.002$ for all final states and \xp bins), suggesting that no additional momentum unfolding is needed. Hence, the extracted signal yields are next corrected for detector acceptance and reconstruction efficiency.  
In the first step, the reconstruction within the barrel region is addressed, where for each studied final-state particle the extracted signal yields in the MC are compared to the true signal yields in the generated MC, still within the kinematic region of the barrel acceptance. In the generated MC, the vertex requirements are dropped. At this stage, also any remaining selection requirements such as the visible-energy, heavy-jet mass, track-multiplicity, and minimum momentum constraints are lifted. As can be seen in Fig.~\ref{fig:acc1}, the efficiencies are generally flat or slightly increasing for intermediate \xp but tend to drop towards lower \xp as well as at \xp close to one. A notable exception is the $K_S^0$ for which the efficiencies decrease monotonically with \xp. This is driven by the vertex requirement, which for long-lived particles becomes more stringent with increasing momentum and thus average laboratory flight distance. Closely related, a lower track reconstruction efficiency for more displaced tracks can also contribute to this behavior. The efficiencies for the different methods of extracting the signal yields are shown as well. They are very similar overall, but can differ at times due to the flexibility of the background functions. They are treated separately in order to account for possible systematic under- or over-estimations due to the functional forms used for the background. 
The efficiencies from the MC sample that additionally includes $\Upsilon(4S)$ decays are also shown for comparison. At higher \xp, which cannot be reached in $\Upsilon(4S)$ decays, the efficiencies are the same. At low \xp, where the $\Upsilon(4S)$ decays dominate, higher efficiencies are seen, particularly for the \(D\) mesons. These differences originate from the different polar-angular distributions of the original particles. \(D\) mesons from charm-pair production still follow the $1+\cos^2 \theta$ behavior of fermionic two-to-two processes (where $\theta$ is the polar angle in the center-of-mass system), while the $\Upsilon(4S)$ resonance is produced at rest and the subsequent decays into charmed mesons result in a
distribution that is flat with respect to $\cos\theta$, or even maximal at central angles. For $D^*$ mesons these differences are not quite as visible. For the lighter particles, any similar differences are hardly discernible.

\begin{figure*}[htb]
    \centering
    \includegraphics[width=0.9\textwidth]{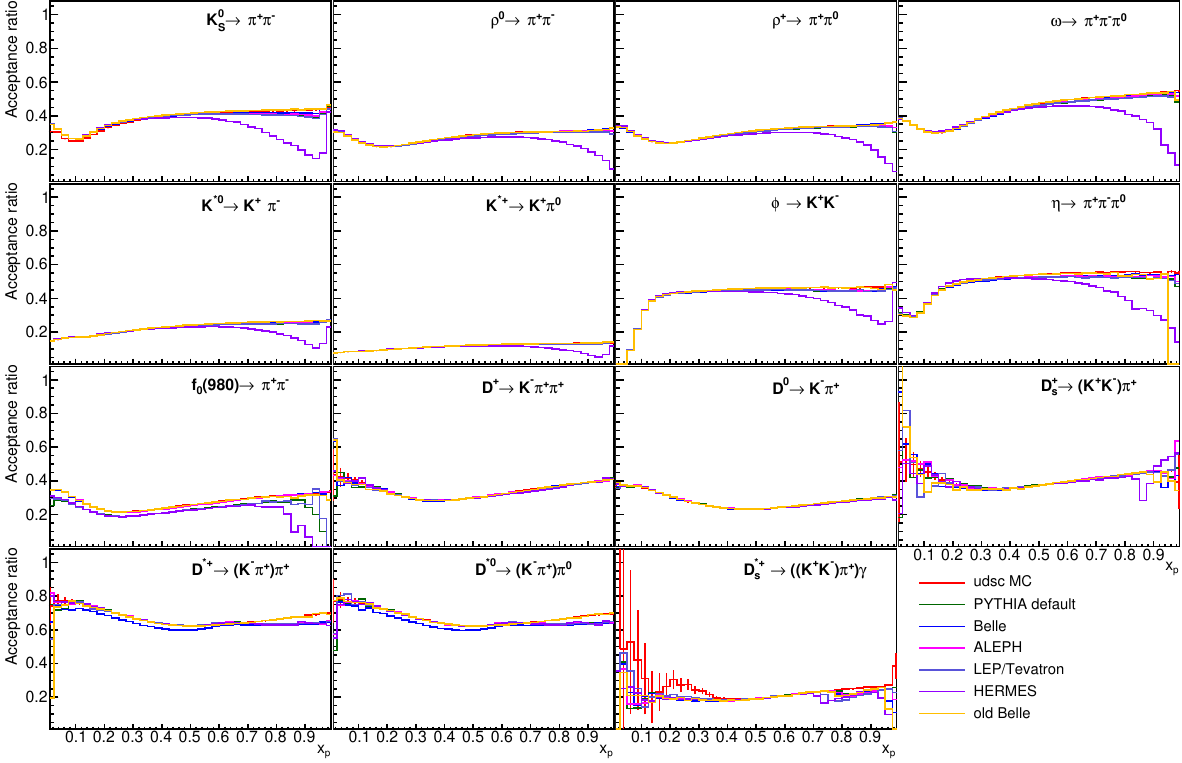}
    \caption{Acceptance efficiencies from the full acceptance to the barrel part of the Belle detector as a function of \xp for the different particles analyzed. The colored lines show the results from the various {\sc pythia} tunes (using udsc samples) described in the text that are used to evaluate the tune-dependent systematic uncertainties.}
    \label{fig:acc2}
\end{figure*}

In the second step of the acceptance correction, the generated MC yields within the barrel part of the detector are compared to the generated yields with $4\pi$ acceptance. The efficiencies due to the tails of the very wide resonances are also addressed in this correction step. The corrections for branching ratios of $K^*$ decays into the charged kaons and a pion, which are the ones considered in this analysis, are handled here via the MC (although those are effectively just the isospin Clebsch-Gordan coefficients for decays into the charged and neutral kaon final-state combinations). 

The extrapolation to $4\pi$ acceptance does depend on the shape of the fragmentation functions, via its implicit polar angular dependence and transverse momentum generation, introducing a tune dependence. 
Therefore, various MC tunes are used to compare their impact on the acceptance correction. As can be seen in Fig.~\ref{fig:acc2}, the efficiencies for nearly all tunes are within a few percent of each other. These variations are considered as a correlated systematic uncertainty. The only tune that deviates significantly is the {\sc hermes} tune, which has failed to describe any of the recent Belle related fragmentation measurements and is thus not considered for inclusion in the systematic uncertainties.  


In the discussion of reconstruction efficiencies, low-momentum particles require special attention. For laboratory momenta of the decay hadrons below 500 MeV, the reconstruction efficiencies are found to be overestimated in the Belle MC, e.g., from comparisons of meson cross sections that included in the analysis chain low-momentum decay particles or not. For the mesons that have a large enough phase space for their decay particles to result in sufficient yields in the two cases of including or excluding low-momentum particles, the comparison can be used to find the optimal low-momentum reconstruction efficiency. 
Three functional forms of the decay particles' laboratory momenta are considered for the additional low-momentum efficiency correction, together with their relative contributions for each \xp bin. A quadratic form provides the best agreement and is thus assigned as the default correction. The variations around the best functional form, as well as the other parameterizations are assigned as additional systematic uncertainties in this analysis. Due to this additional increase in systematic uncertainties, the low-momentum particle selection is only used when required by phase space. Consequently, for lighter hadrons this selection is applied for $\xp < 0.25$ ($0.35$) for final states of two (three) particles. For regular $D$ mesons and $D_s$, the high-momentum selection is sufficient, but for $D^*$ the low-momentum selection is needed for \xp below 0.9 due to the small mass difference between the $D^*$ and the $D^0$. 

\subsection{ISR correction}
\begin{figure*}[htb]
    \centering
    \includegraphics[width=0.9\textwidth]{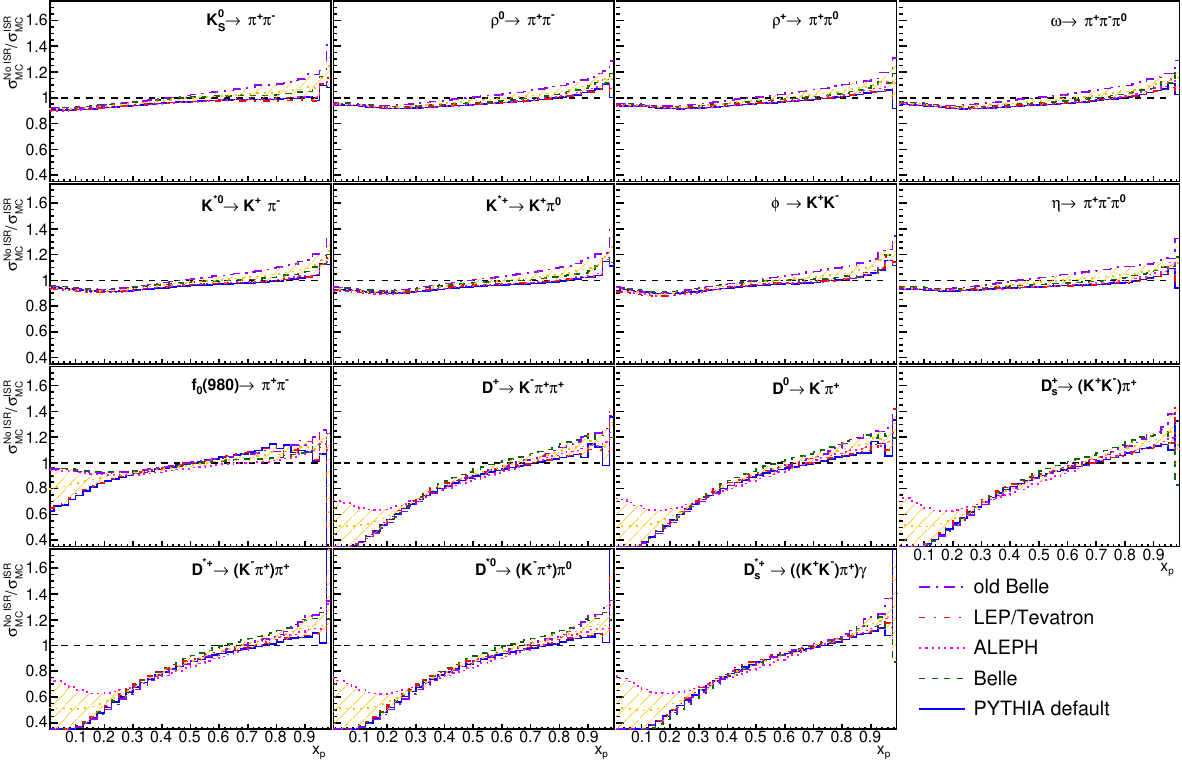}
    \caption{ISR correction ratios as a function of \xp as evaluated by the ratio of MC-simulated yields with ISR switched off to MC simulations including ISR. To evaluate the tune dependence, various MC tunes are compared. The overall variation between them is displayed as the yellow dashed area and enters into the systematic uncertainty calculation as described in the
text.}
    \label{fig:isr}
\end{figure*}

Similar to other recent Belle fragmentation measurements, the impact of initial-state photon radiation (ISR) is addressed by comparing generated MC signal yields including ISR to those where it is switched off. 
As the ISR photon takes away energy from the quark-antiquark system, the phase space for high-\xp hadron production (relative to the nominally available maximum momentum) gets reduced and the ratio between no-ISR and ISR becomes larger than unity. By contrast, the fraction of low-\xp hadrons increases and hence the ratio here becomes smaller than unity. This effect appears to become more prominent for the heavier particles. For lighter mesons, the ISR ratios vary only up to ten percent around unity, while for $D$ mesons, the ratios fall as low as 0.4 at low \xp and reach nearly 1.4 at high \xp, as can be seen in Fig.~\ref{fig:isr}. 
Different MC tunes (cf.~Table \ref{tab:jetsettable}) are investigated and show a slightly different magnitude of the ISR effects, with typically the {\sc aleph} or old Belle tunes deviating most from the other tunes. These tune variations are again assigned as systematic uncertainties.

\section{Systematic tests\label{sec:systematics}}
\subsection{Consistency tests}
\begin{figure*}[htb]
    \centering
    \includegraphics[width=0.9\textwidth]{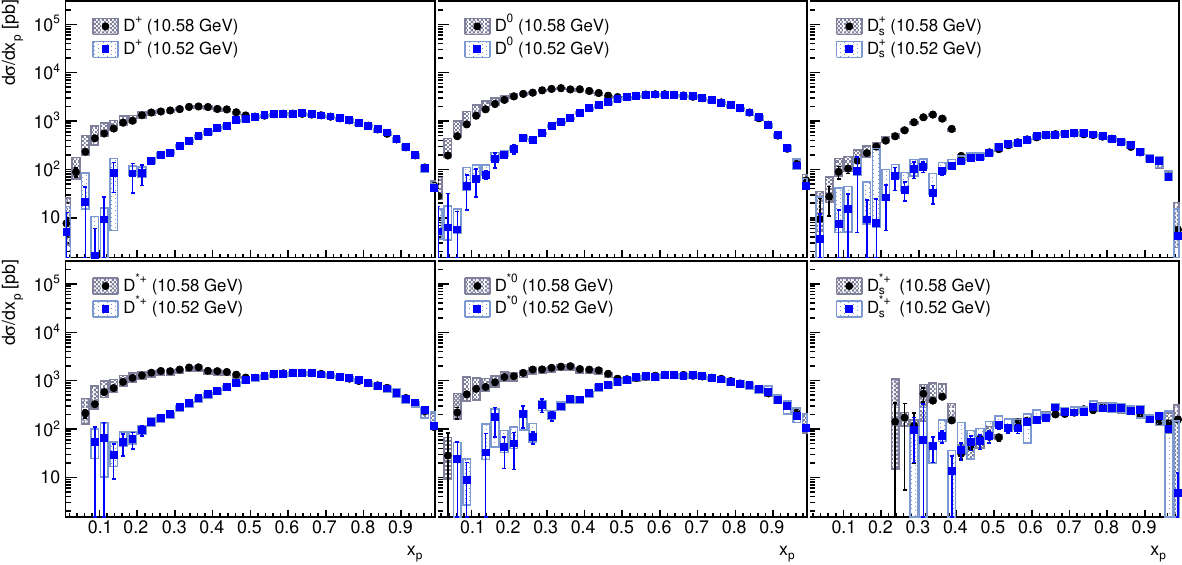}
    \caption{Production cross sections as a function of \xp for $D^+$, $D^0$, $D_s^+$, $D^{*+}$, $D^{*0}$, and $D^*_s$ for continuum (blue) and on-resonance (black) data. The low-\xp point selection is as described in the text.}
    \label{fig:dmcontres}
\end{figure*}

\begin{figure*}[htb]
    \centering
    \includegraphics[width=0.9\textwidth]{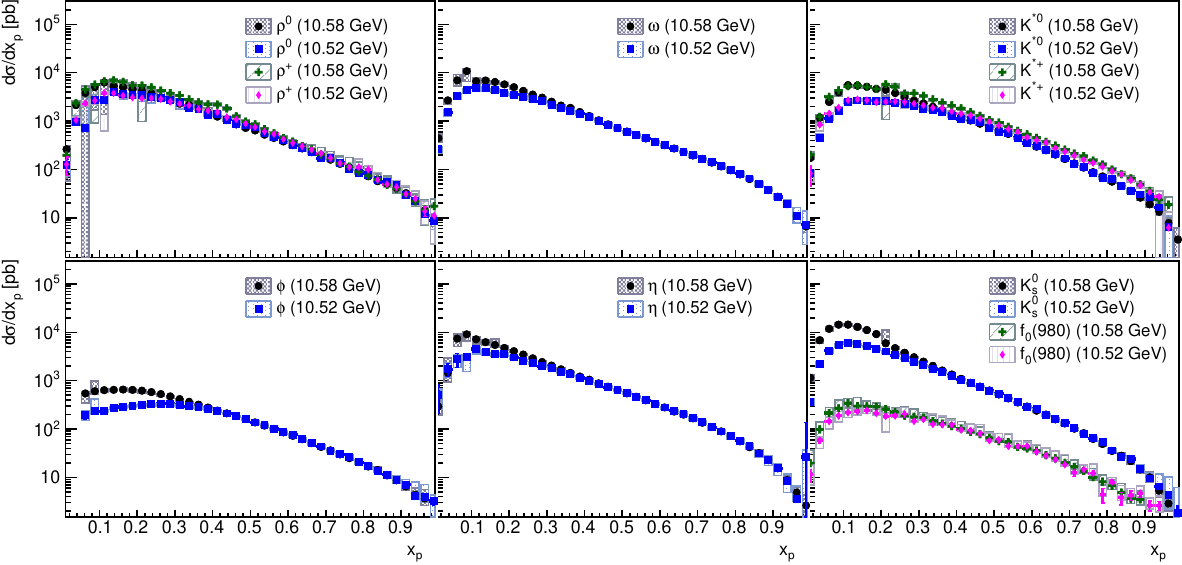}
    \caption{Production cross sections as a function of \xp for $\rho^+$, $\rho^0$, $\omega$, $K^{*0}$, $K^{*+}$, $\phi$, $\eta$, $K_S^0$, and $f_0(980)$ for continuum (blue and pink points) and on-resonance (black and green points) data. }
    \label{fig:vmcontres}
\end{figure*}

The \(\Upsilon(4S)\) dominantly decays into a pair of \(B\) mesons ($>$96\% \cite{ParticleDataGroup:2024cfk}). While the decay products from those can contribute at \xp of around 0.5 and below, they cannot contribute at higher \xp, given that the $\Upsilon(4S)$ and its subsequent \(B\) mesons decay nearly at rest. As such, the continuum and on-resonance data samples should be consistent for the higher \xp region. The comparison can be seen in the differential cross sections in Fig.~\ref{fig:dmcontres} for the $D$ mesons and in Fig.~\ref{fig:vmcontres} for the lighter mesons. In particular for the $D$ mesons, the different peaking structures from fragmentation at high \xp and the peaks from the $\Upsilon(4S)$ resonance and subsequent decays are clearly distinguishable. In the higher \xp region, the production cross sections are consistent with each other, as expected. 
For the lighter mesons, the distinction between the two data samples is not as clearly visible as the peak in the fragmentation functions moves to much lower \xp \cite{Albino:2004xa}. Nevertheless, one can see the agreement at higher \xp, while at lower \xp differences due to the $\Upsilon(4S)$ decays are visible. Given the substantially larger statistical precision of the on-resonance data, which also results in reduced fluctuations in the fitted values, the final  cross sections presented here use the on-resonance data for \xp above 0.55 while in the region below, the continuum data is used. 

Another systematic consistency test is the comparison of the cross sections obtained from different final states for the same particle. The $D^0$-meson reconstruction via its $K\pi$ decay, considered here, can be compared to its reconstruction via $K\pi\pi^0$ decay as well as its two Cabibbo-suppressed $\pi\pi$ and $KK$ decays. 
The cross sections for all four final states 
are found to be consistent with each other. 

For the $D^+_s$ meson, the decay considered here into $\phi (\rightarrow KK) \pi$ can be compared to the $K^0_S (\rightarrow \pi\pi)K$ decay. The final state of three pions was not feasible to study due to a large background. The results from these different decay modes are again found to be consistent with each other within uncertainties, thus not requiring assignment of any additional systematic uncertainty.

As an additional consistency test, the neutral pions that contribute in several final states, are compared to the previously published charged-pion cross sections and found to be consistent. 

\subsection{Overall systematic uncertainties}

\begin{figure*}[htb]
    \centering
    \includegraphics[width=0.9\textwidth]{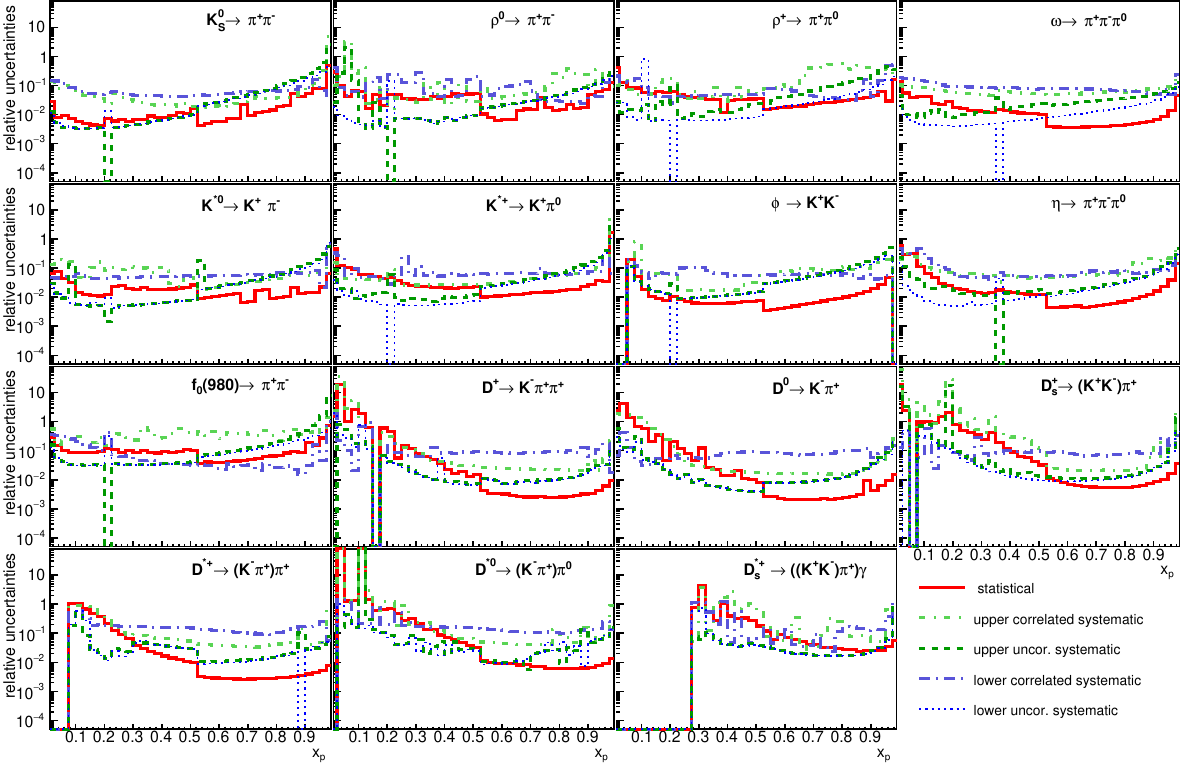}
    \caption{Uncertainty budgets normalized to the cross sections are shown for continuum (on-resonance) data below (above) \xp $=0.55$ as a function of \xp. The uncertainties are displayed for statistical uncertainties (red lines), correlated upper and lower systematic uncertainties (green and blue dash-dotted lines), and uncorrelated upper and lower uncertainties (green dashed and blue dotted lines).}
    \label{fig:syst}
\end{figure*}

The systematic uncertainties that are previously mentioned in the different correction steps are combined to provide the final values. They are summarized in Fig.~\ref{fig:syst} for all relevant hadrons as a function of \xp after merging the continuum and on-resonance data sets for the regions below and above \xp of 0.55, respectively (notably visible in the drop of statistical uncertainties), and the high- and low-momentum selections where necessary. For simplicity, only the quadratic sum of all upper and lower correlated and likewise uncorrelated uncertainties are shown relative to the actual cross sections. 
All systematic-uncertainty contributions are separately tabulated in the supplement \cite{supplement}, together with the cross sections. As can be seen, the measurements for all hadrons are systematics dominated. Among the systematic uncertainties, the correlated uncertainties generally dominate over the uncorrelated ones. The tune-dependent uncertainties are generally the largest, and particle-identification corrections are significant at low \xp. For $D^*$, and at lower \xp for lighter hadrons, the uncertainties due to the additional low-momentum efficiency corrections also contribute significantly. 

A total scale uncertainty of 1.4\% for the luminosity normalization and an uncertainty of 0.35\% per charged decay particle from track reconstruction efficiencies is assigned. An uncertainty of 1.7\% is assigned for the neutral pion reconstruction. For momenta below 200 MeV the uncertainties of 1.4\% and 2.4\% for charged particles and neutral pions, respectively, are assigned according to their relative contributions in signal MC. All these uncertainties are reflected in the total systematic uncertainty budget, as shown in Fig.~\ref{fig:syst}. Overall scale uncertainties arising from the limited knowledge of the branching fractions involved (cf. Table \ref{tab:BRs}) are kept separate to facilitate updating the results with newer branching fractions when available.

\section{Results\label{sec:results}}
The final results are obtained after applying all the correction steps to the fitted yields after subtracting non-$q\bar{q}$ backgrounds, namely the normalization with accumulated luminosities $\mathcal{L}$ and branching fractions $\mathcal{B}$, \xp bin widths, corrections for the acceptance and reconstruction efficiencies ($\epsilon_{\mathrm{acc}}, \epsilon_{\mathrm{rec}}$), additional low-momentum efficiency corrections ($\epsilon_{\mathrm{Lowp}}$), and the corrections for the initial-state radiation ($\epsilon_{\mathrm{ISR}}$) as described in the previous sections:
\begin{equation}
    \frac{d \sigma}{d\xp} = \frac{N_{\mathrm{Fit}} - N_{\mathrm{nonq\bar{q}}} }{\mathcal{L} \times \mathcal{B} \times  \Delta \xp}\frac{1}{ \epsilon_{\mathrm{rec}} \times \epsilon_{\mathrm{acc}} \times \epsilon_{\mathrm{Lowp}} \times  \epsilon_{\mathrm{ISR}}}\quad .
\end{equation}
For the acceptance and ISR corrections, the Belle tune was chosen as the central value, while the cross sections obtained from using the other tunes are tabulated in the supplemental material \cite{supplement} and are part of the systematic uncertainties.  

\begin{figure*}[htb]
    \centering
    \includegraphics[width=0.9\textwidth]{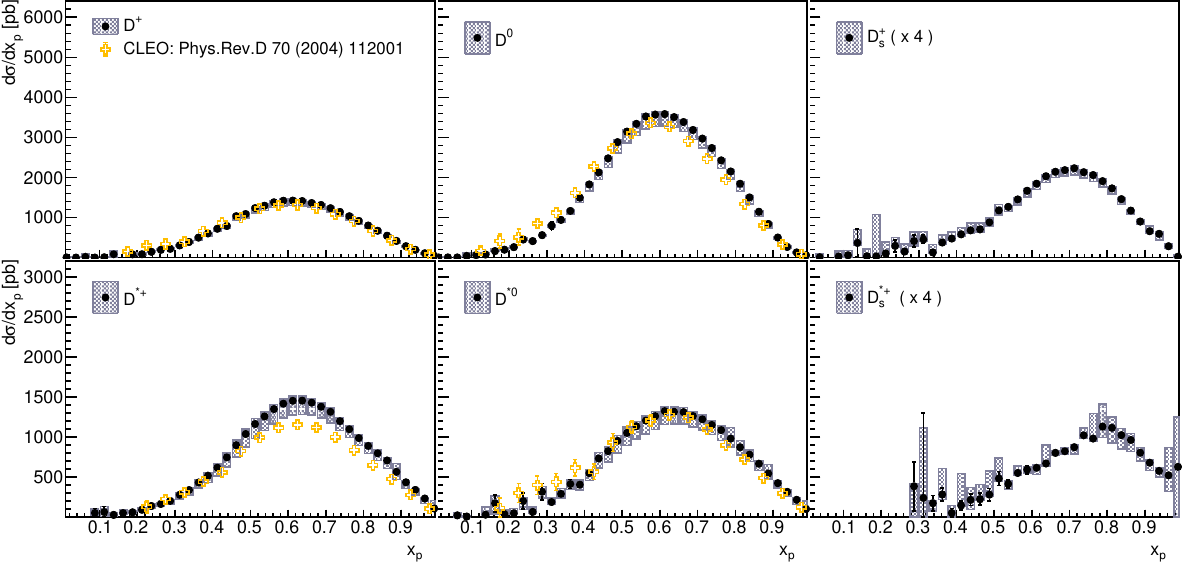}
    \caption{Production cross sections as a function of \xp for $D^+$, $D^0$, $D_s^+$, $D^{*+}$, $D^{*0}$, and $D^*_s$. The $D_s$ results are scaled by a factor four for better visibility. For comparison, also the previously published results by CLEO \cite{CLEO:2004enr} are displayed as yellow crosses, where available. The low-\xp point selection is as described in the text. }
    \label{fig:dmfinal}
\end{figure*}

As apparent from Fig.~\ref{fig:dmfinal}, the charm-meson results are in good agreement with the older results from CLEO \cite{CLEO:2004enr}, particularly at larger values of \xp, with the exception of the $D^{*+}$ which are slightly larger. They are also consistent with the previous Belle results \cite{Belle:2005mtx} (not shown), which they supersede. In both cases, the shapes show a smaller tail than the older results at low \xp and a slightly larger tail at high \xp. These differences originate from the ISR correction, which was
taken to be flat in the older publications, but found to have a distinct shape here. Before that correction, also the shapes of the old and new Belle cross sections agree quite well with each other. BaBar has published results on $D_s$ meson production \cite{BaBar:2002ncl}. The integrated values for the production cross sections are consistent with the ones from Belle (taking into account the changes in the BR involved reported by the PDG between then and now). However, a comparison of the differential cross sections is hampered by unclear figure legends and an apparent lack of tabulated values.


\begin{figure*}[htb]
    \centering
    \includegraphics[width=0.9\textwidth]{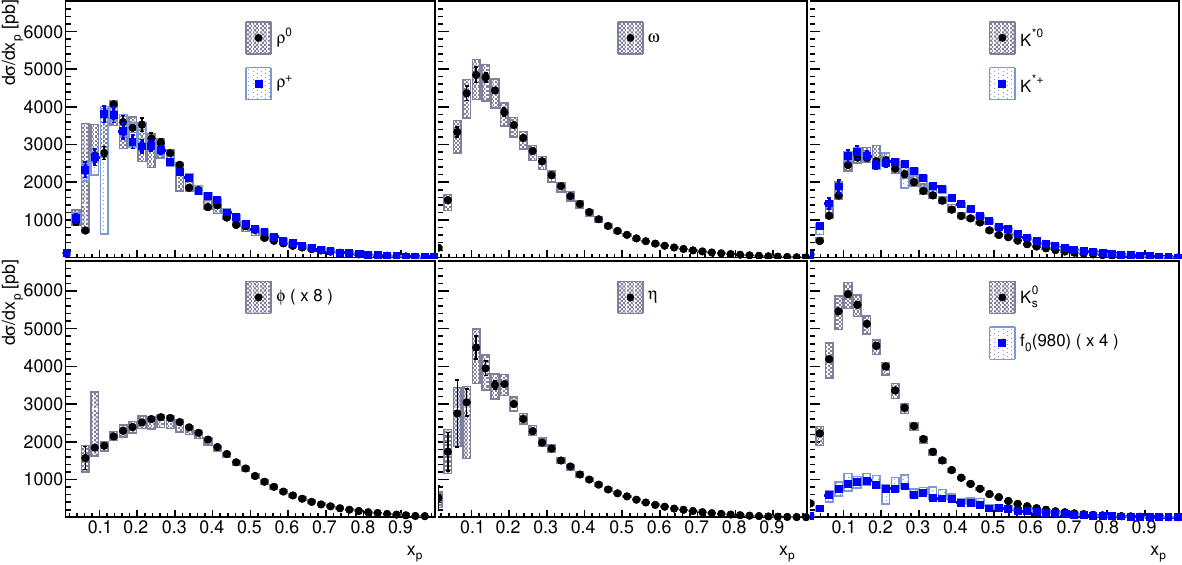}
    \caption{Production cross sections as a function of \xp for $\rho^+$, $\rho^0$, $\omega$, $K^{*+}$, $K^{*0}$, $\phi$, $\eta$, $K_S^0$, and $f_0(980)$. The $\phi$ and $f_0(980)$ cross sections are scaled by a factor eight and four, respectively, for better visibility.}
    \label{fig:vmcxsec}
\end{figure*}

For the light vector mesons, no previous measurements at this energy exist. Together with selected other light mesons, their production cross sections are shown in Fig.~\ref{fig:vmcxsec} as a function of the fractional momentum. As expected from local hadron parton duality \cite{Azimov:1985by,Dokshitzer:1991eq}, the peak of each distribution increases with the mass of the particle. For the lightest particles, the peak is at \xp around 0.1 and increases up to 0.2 to 0.3 for heavier non-charmed particles. For $D$ mesons it is substantially higher. 

\subsection{Isospin symmetry comparisons}
\begin{figure*}[htb]
    \centering
    \includegraphics[width=0.48\textwidth]{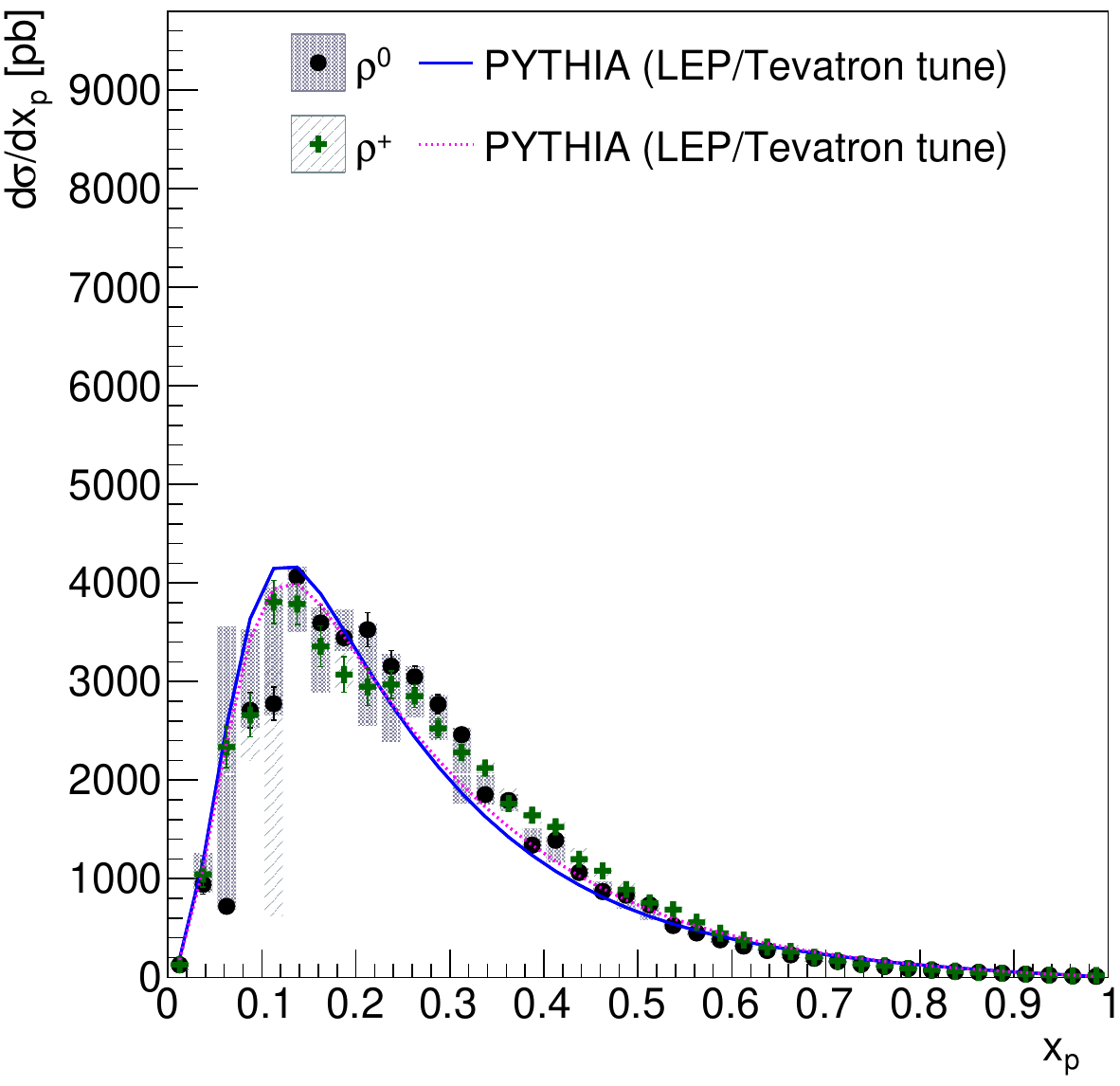}
    \includegraphics[width=0.48\textwidth]{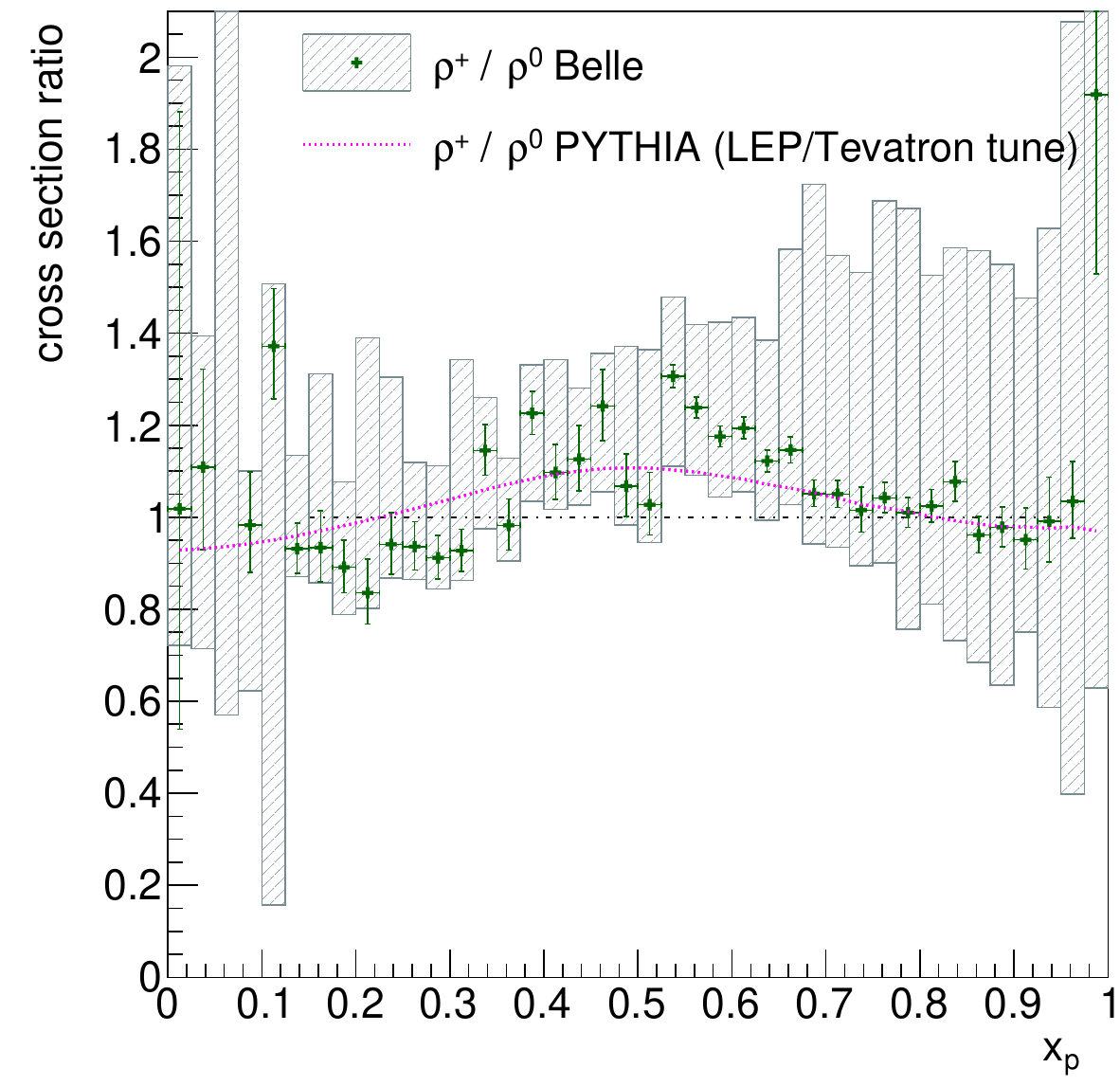}
    \caption{Left: Comparison of neutral and charged $\rho$ meson cross sections for both the Belle data and a {\sc pythia} MC simulation using the {\sc lep}/Tevatron fragmentation tune. Right: Corresponding ratios of charged to neutral $\rho$ cross sections, both for Belle data and MC. }
    \label{fig:rhoratios}
\end{figure*}

Isospin symmetry should mostly hold for the FFs of the three charge-states of the $\rho$ meson. When comparing the neutral and both charged $\rho$ mesons, this can be confirmed for higher \xp, however at intermediate \xp an excess of up to 20\% for the charged mesons is observed. In simulations, a small excess is seen due to the decays from charm production, but only at the level of a few percent maximally. A comparison of both charged and neutral $\rho$ mesons is displayed in Fig.~\ref{fig:rhoratios}.

\begin{figure*}[htb]
    \centering
    \includegraphics[width=0.48\textwidth]{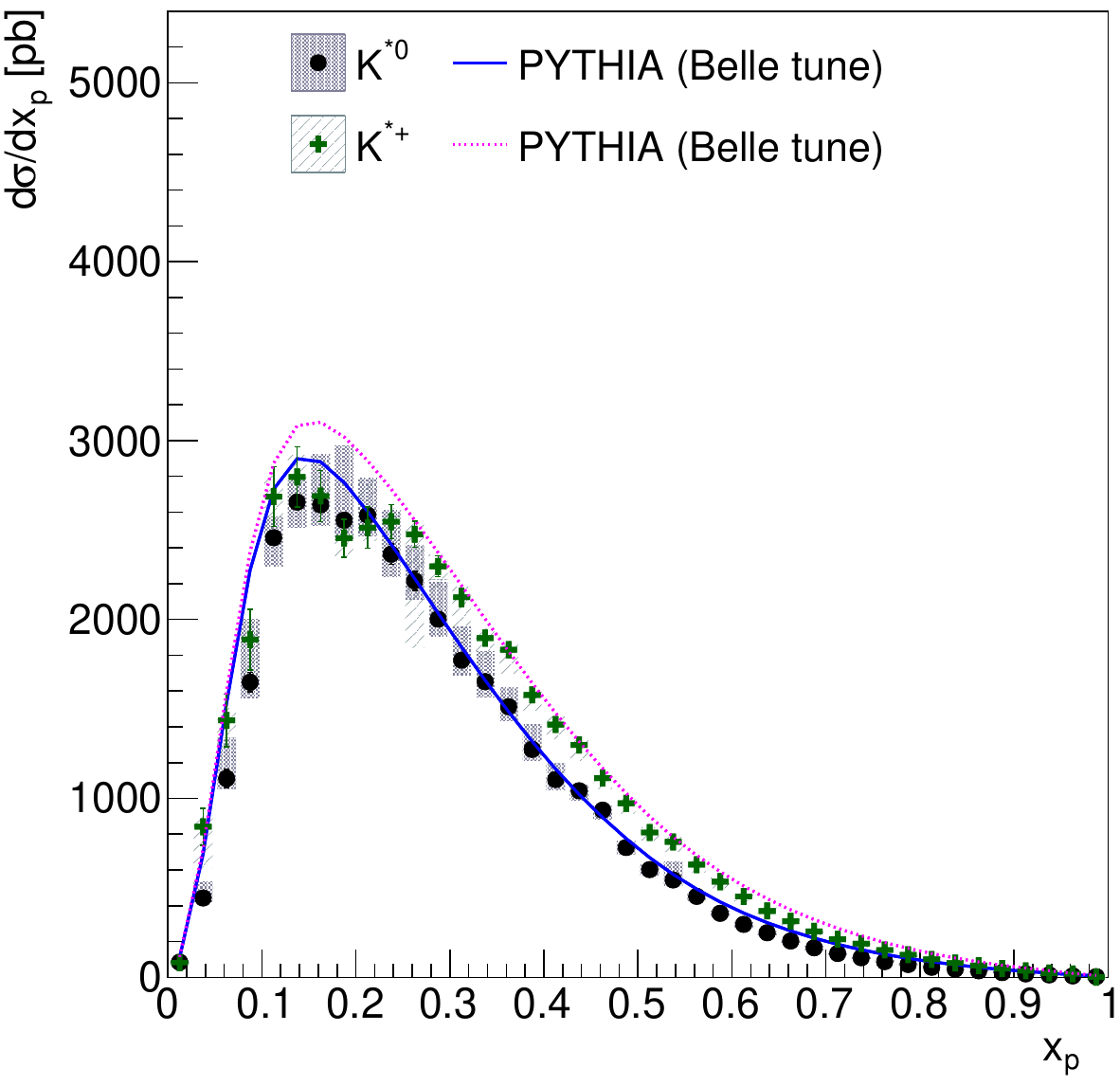}
    \includegraphics[width=0.48\textwidth]{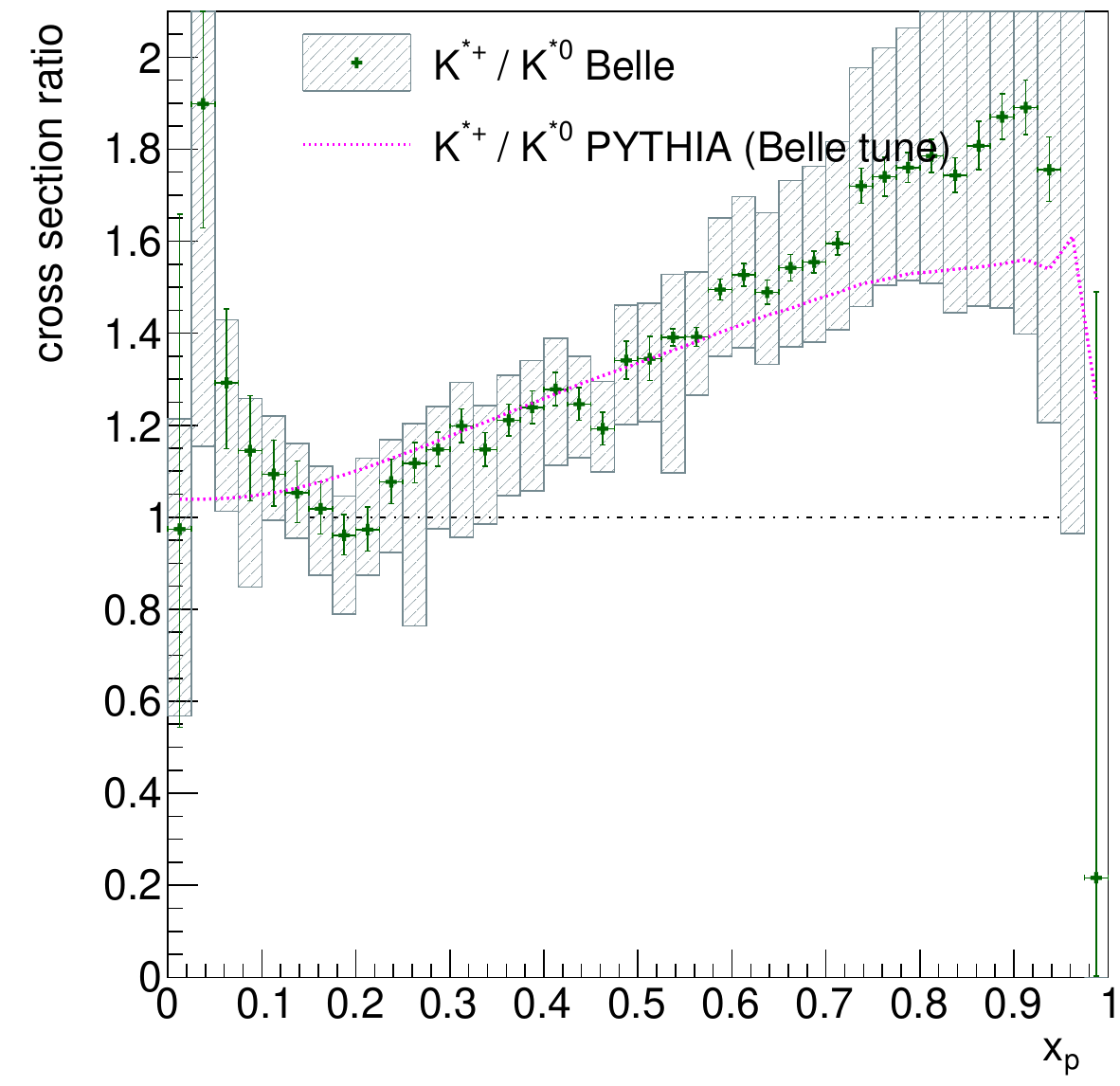}
    \caption{Left: Comparison of neutral and charged $K^*$ meson cross sections for both the Belle data and a {\sc pythia} MC simulation using the Belle fragmentation tune. Right: Corresponding ratios of charged to neutral $K^*$ cross sections, both for Belle data and MC.}
    \label{fig:kstarratios}
\end{figure*}

For $K^*$, the fragmentation functions should also be isospin symmetric. However, in the initial process during $e^+e^-$ annihilation, due to the charge factor, $u\bar{u}$ and $d\bar{d}$ pairs are not produced equally, resulting in a difference in the cross sections. This is not the case for pions or $\rho$ mesons as the initially produced quark-antiquark pair always provides a matching valence-parton. In the MC, the fragmentation from strange quarks appears to be equal for both the neutral and charged kaons, but the differences in initial up and down quarks result in an enhancement of charged $K^*$ production that is increasing with \xp. This increase at higher \xp is expected as the favored fragmentation (where the fragmenting quark is a valence particle in the detected hadron) from the initial up or down quark is larger there, as discussed for example in Ref.~\cite{deFlorian:2017lwf}. This is also observed in the data where the cross sections are comparable at low \xp but an excess of charged $K^*$ is visible and increasing with \xp, as expected. These comparisons are displayed in Fig.~\ref{fig:kstarratios}.

\begin{figure*}[htb]
    \centering
    \includegraphics[width=0.48\textwidth]{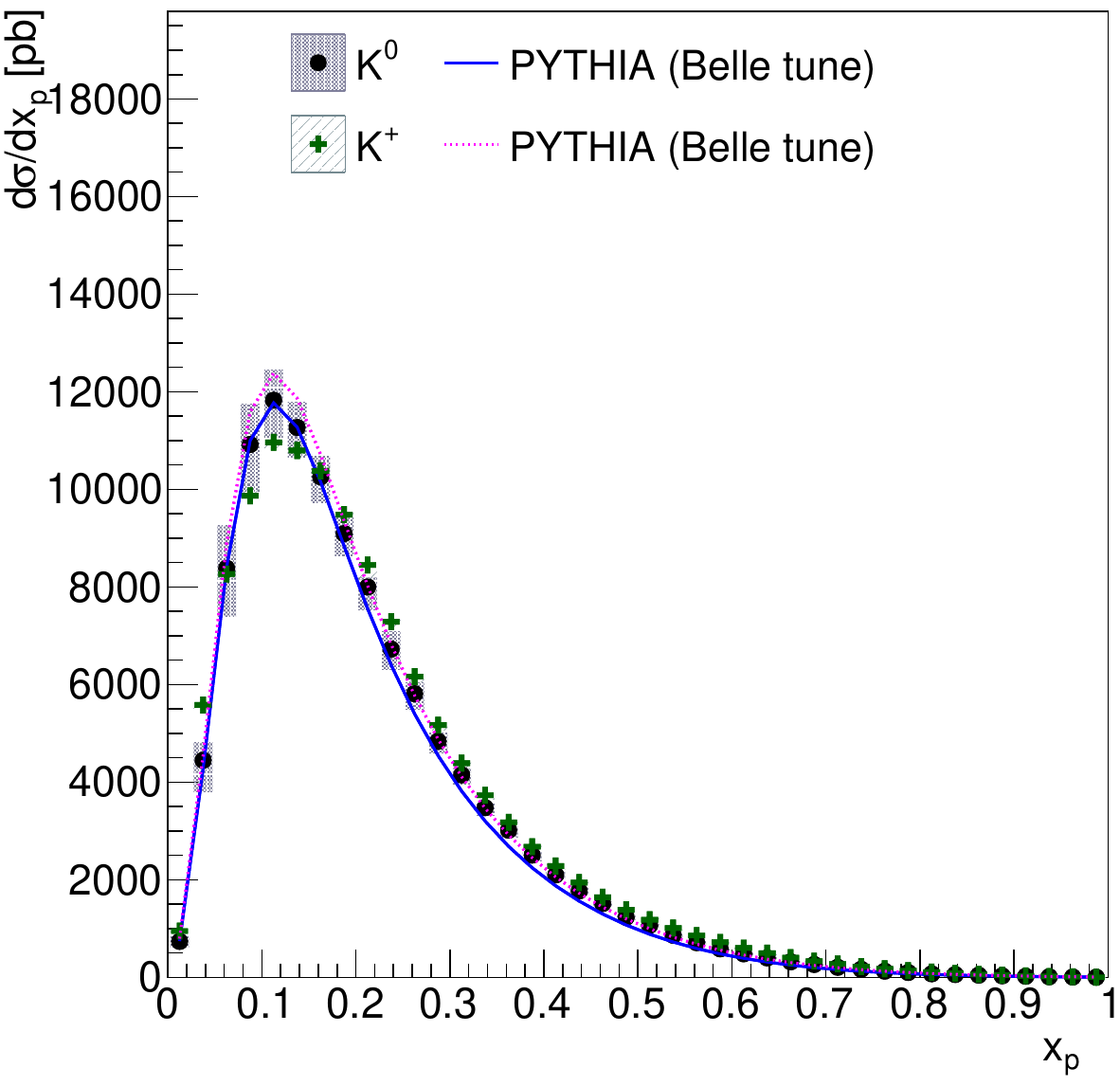}
    \includegraphics[width=0.48\textwidth]{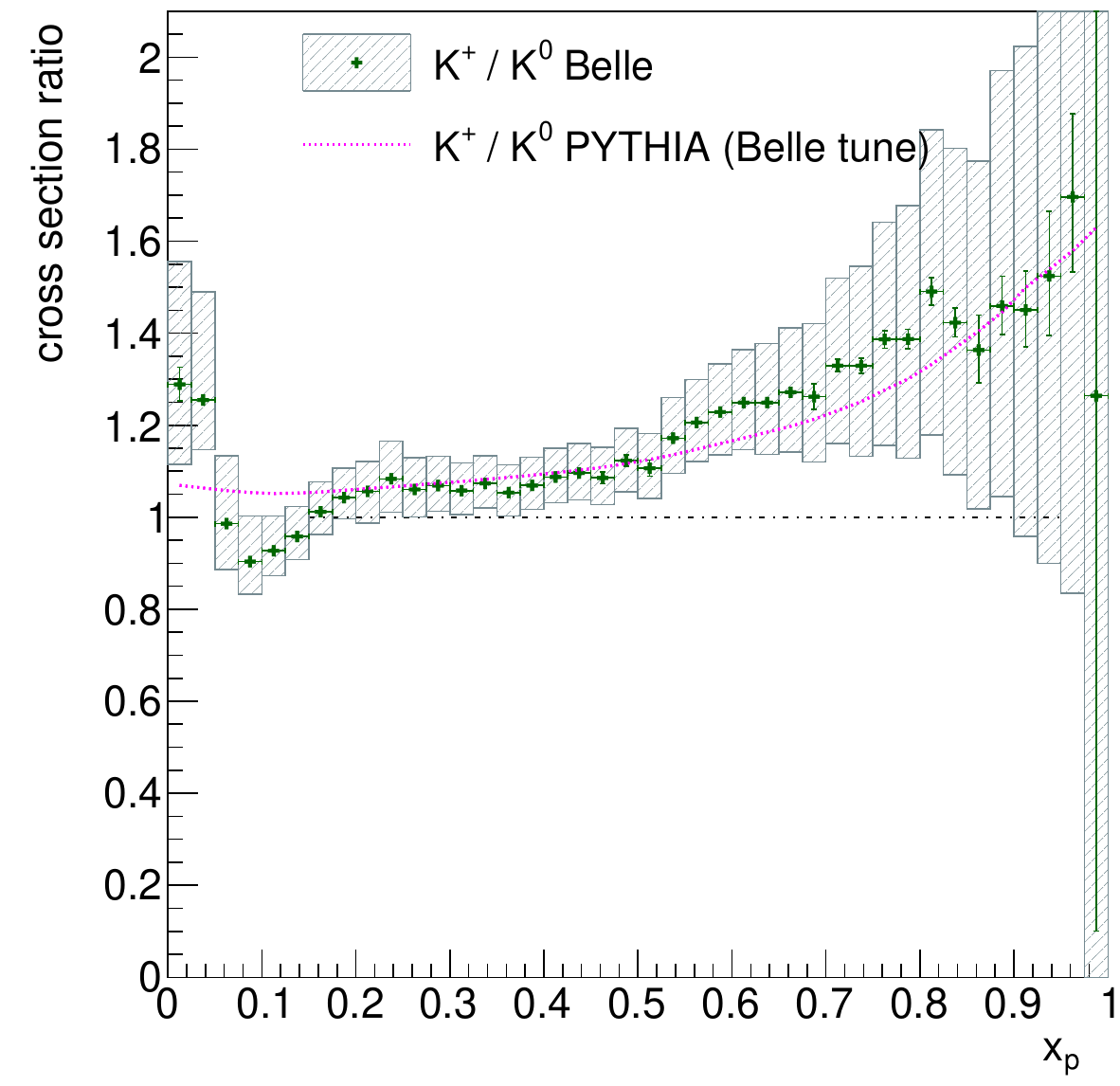}
    \caption{Left: Comparison of neutral and charged $K$ meson cross sections for both the Belle data and a {\sc pythia} MC simulation using the Belle fragmentation tune. Right: Corresponding ratios of charged to neutral $K$ cross sections, both for Belle data and MC.}
    \label{fig:kaonratios}
\end{figure*}

A similar, but slightly less pronounced behavior is also seen for the pseudoscalar kaons, where the differences amount to about 5\% at intermediate \xp and are also consistent with MC expectations. Those are shown in Fig.~\ref{fig:kaonratios}.

\subsection{Direct vector vs pseudoscalar fragmentation studies}
\begin{figure*}[htb]
    \centering
    \includegraphics[width=0.9\textwidth]{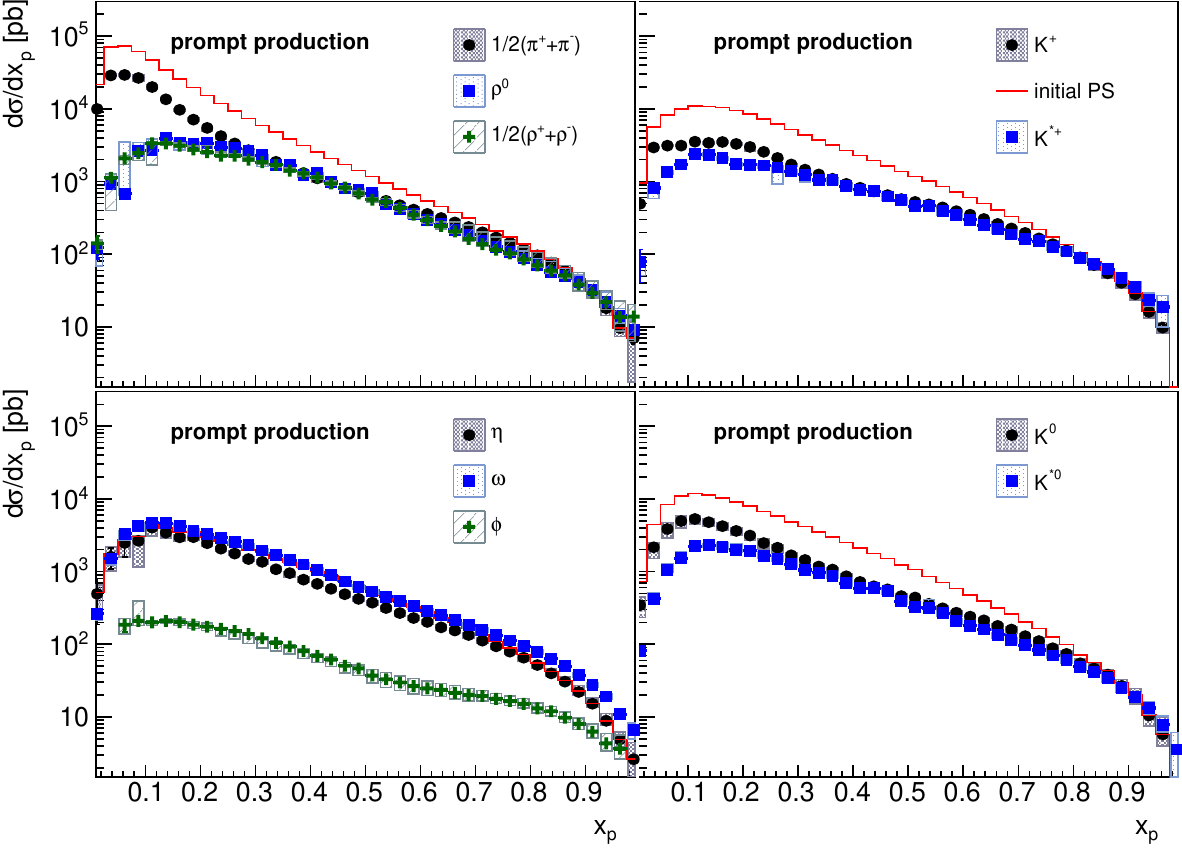}
    \caption{Comparison of \xp differential prompt production rates for pseudoscalar mesons to those for vector-meson cross sections after subtracting the contributions from decays of the parent-particles analyzed in this work into these final particles. The red histograms indicate the cross sections of the corresponding pseudoscalar mesons before subtraction. Top left: Comparison of charge-averaged charged pions (black points) to charge-averaged charged $\rho$ mesons (dark green, upward triangles), and $\rho^0$ (blue squares). Top right: Comparison of charged kaons (black points) to $K^{*+}$ (blue squares). Bottom left: $\omega$ (black points), $\eta$ (blue squares), and $\phi$ (dark green, upward triangles). Bottom right: $K^0$ (black points) and $K^{*0}$ (blue squares).}
    \label{fig:psvm}
\end{figure*}

For the lighter mesons, it is of interest to compare the differential cross sections of {\em promptly} produced light pseudoscalar and vector mesons. Cross sections for $\rho$, $\omega$, $K^*$, and $\phi$ are shown in Fig.~\ref{fig:psvm} in comparison to $\eta$, $K^0$ as well as charged pions and kaons. The charged pions and kaons are extracted here in an \xp binning while the previously published cross sections \cite{Belle:2020pvy} were binned in the energy fraction. For these {\it direct} production rates, the feed-down contributions from all the final states considered in this analysis are subtracted, based on their measured cross sections and MC-based \xp migration matrices between parent and daughter particles. It should be noted that such feed-down subtracted cross sections do not fulfill the formal definitions of fragmentation functions, which are inclusive objects including all strong decays. They nevertheless indicate the differences in direct fragmentation, which may be relevant to fragmentation settings in MC event generators. 

To compare similar quantities, the average of positive and negative pions is taken to compare to the $\rho^+$ and $\rho^0$ mesons. It can be seen that at high \xp, pions and the $\rho$ vector mesons have more similar cross sections, suggesting that light pseudoscalar and vector mesons are produced about equally in fragmentation. At lower \xp, the decays from other particles, which are not accounted for here, contribute to pions and thus their cross sections start exceeding those of the vector mesons significantly. Based on MC studies, the fraction of unaccounted for contributions is below 60\% of the actual directly produced yields for pions, while it is much lower for $\rho$ mesons. However, even in the MC the missing feed-down does not appear to be the main cause of the differences, suggesting that the fairly large mass difference between pions and $\rho$ mesons is more relevant at low \xp for fragmentation.   

For kaons, the comparison between pseudoscalar and vector mesons shows a similar behavior. The charged mesons are again on a similar level at intermediate \xp. The pseudoscalar meson cross sections start exceeding the vector mesons again at lower \xp due to the additional decays, but to a lesser extent than for pions. The MC-based fractions of unaccounted for contributions are smaller here, reaching  up to 25\% for pseudoscalar kaons and even less for other particles. For a consistent comparison of the neutral kaons, all $K^0$ are displayed, based on scaling the $K_S^0$ measurements by a factor of two to include the unaccounted $K_L^0$ contribution.   
The $\eta$ mesons appear to be generally suppressed compared to $\omega$ mesons, but their shapes are similar. As the contribution from $\eta'$ decay is not subtracted here, the direct fragmentation into $\eta$ is even lower (possibly by a factor of more than two, according to MC simulations).
 The $\phi$ mesons are further suppressed, as expected due to the additional strangeness that needs to be created in the fragmentation process.

\begin{figure*}[htb]
    \centering
    \includegraphics[width=0.9\textwidth]{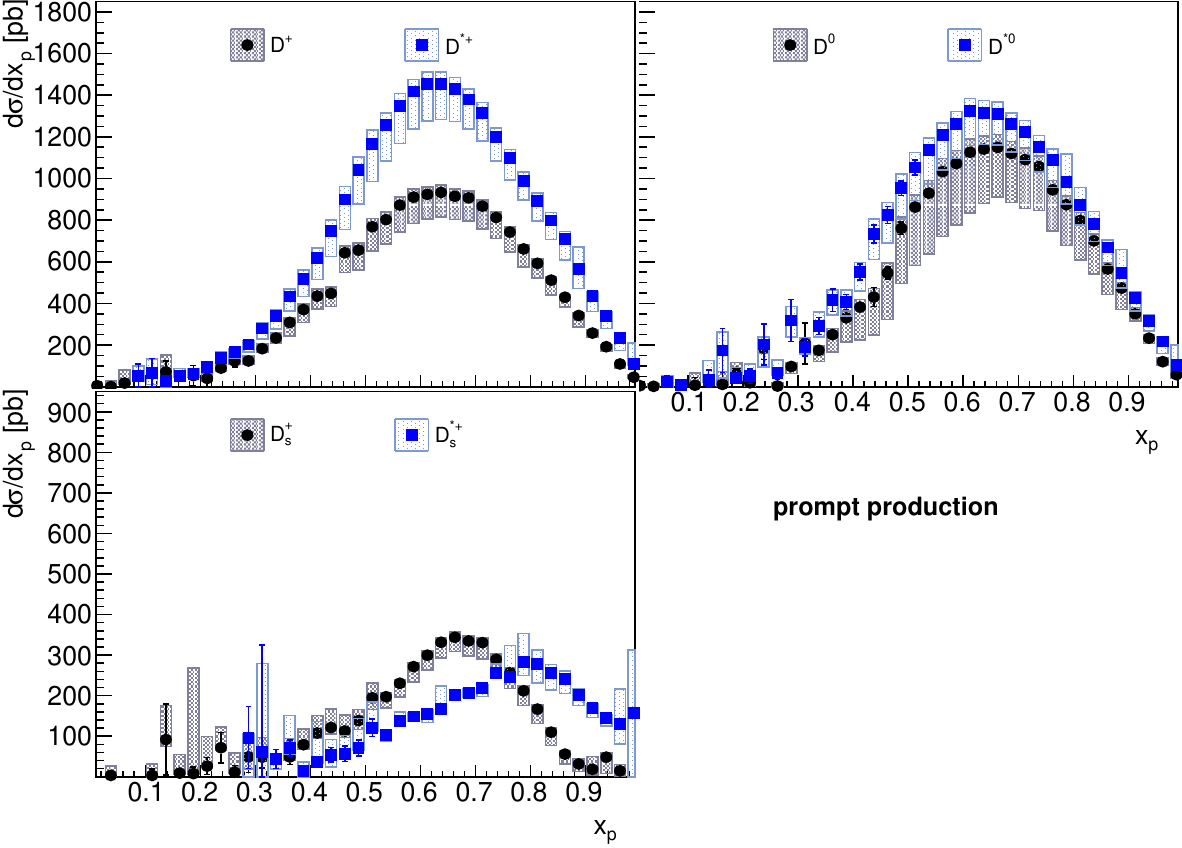}
    \caption{Comparison of pseudoscalar meson cross sections to vector-meson cross sections for various charmed mesons after subtracting the contributions from all decays of the parent-particles analyzed in this study.}
    \label{fig:dmpsvm}
\end{figure*}

Similar comparisons between pseudoscalar and vector mesons are presented in Fig.~\ref{fig:dmpsvm} for the charmed mesons. Also in this case, the feed-down from the higher-mass vector mesons analyzed here have been subtracted, with nearly all $D^*$ mesons decaying into pseudoscalar $D$ mesons. An increased fragmentation for charmed vector compared to pseudoscalar mesons is clearly visible. This is consistent with what most MC generators generally have as their default settings for charm fragmentation. In the case of $D_s^+$, while the overall size is similar to that of the \(D_s^{*+}\), a pronounced difference in the \xp shape is visible with the $D_s^{*+}$ peaking at substantially larger \xp than the $D_s^+$ cross section. 

Additionally, isospin symmetry appears to be roughly fulfilled for both $D^*$ as well as pseudoscalar $D$ mesons, despite the initial cross sections before feed-down correction being significantly larger for neutral $D$ mesons. The latter appear slightly larger but are within the uncertainties that mostly originate from the parent $D^*$ uncertainties. A similar observation was already made in a previous publication \cite{Belle:2005mtx}. 

\begin{figure*}[htb]
    \centering
    \includegraphics[width=0.9\textwidth]{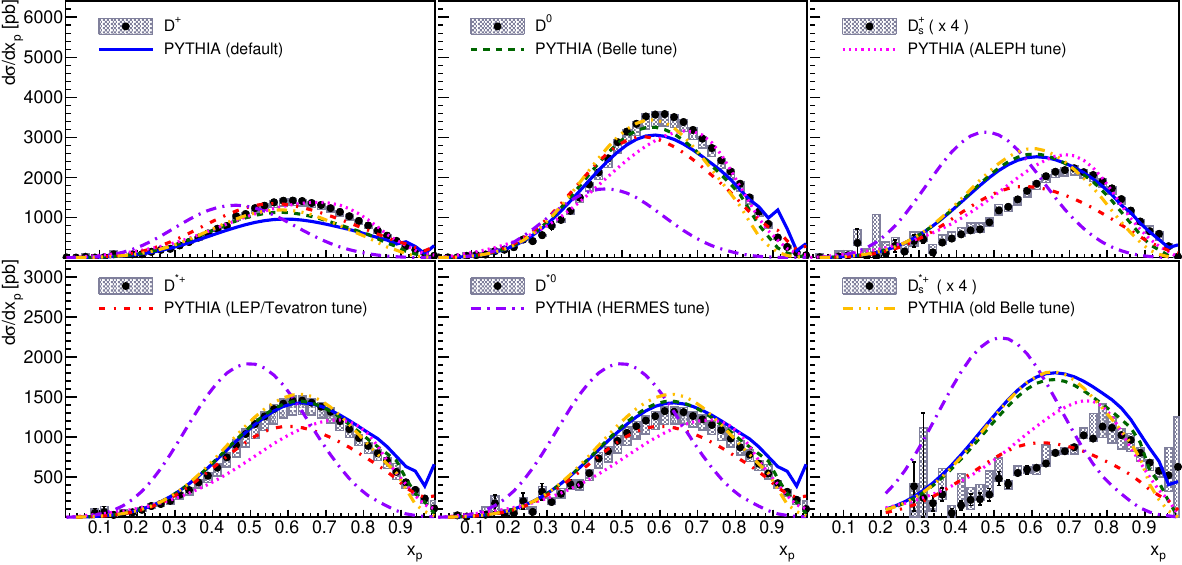}
    \caption{Production cross sections as a function of \xp for $D^+$, $D^0$, $D_s^+$, $D^{*+}$, $D^{*0}$, and $D^{*+}_s$ shown for data (black points), compared to various {\sc pythia} tunes as described in the text. The $D_s$ cross sections are scaled by a factor four. The low-\xp point selection is as described in the text. }
    \label{fig:dmtunes}
\end{figure*}

\begin{figure*}[htb]
    \centering
    \includegraphics[width=0.9\textwidth]{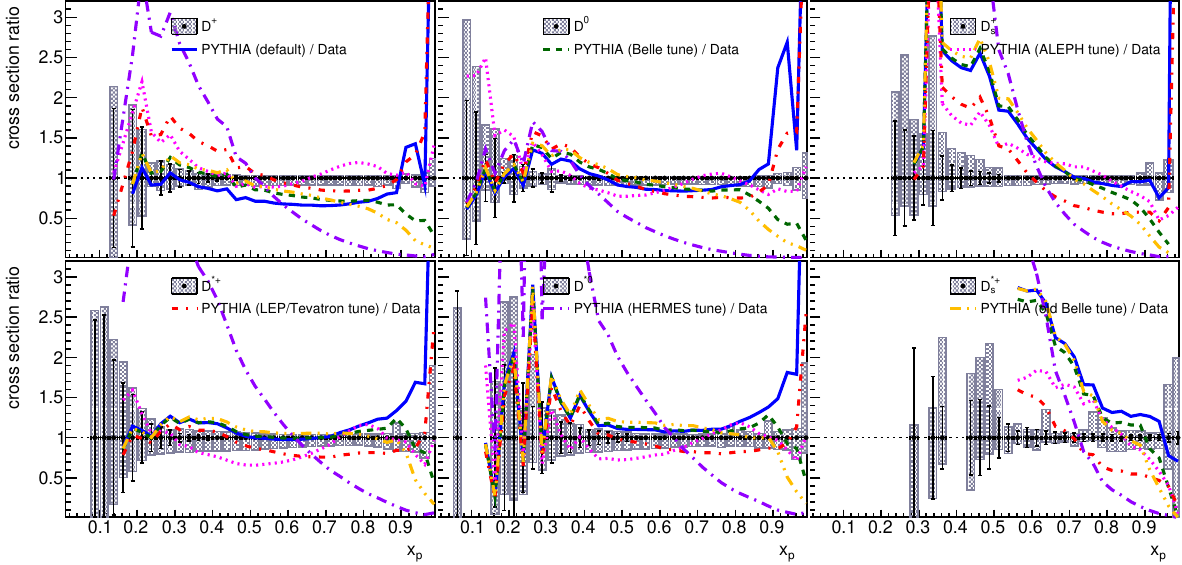}
    \caption{Production cross sections for $D^+$, $D^0$, $D_s^+$, $D^{*+}$, $D^{*0}$, and $D^{*+}_s$ from MC simulations using various fragmentation tunes (as indicated) relative to Belle data as a function of \xp. The data points and uncertainty boxes display the relative uncertainties of the data, whose values are suppressed
          for clarity when exceeding the scale of the plot. }
    \label{fig:dmratios}
\end{figure*}

\begin{figure*}[htb]
    \centering
    \includegraphics[width=0.9\textwidth]{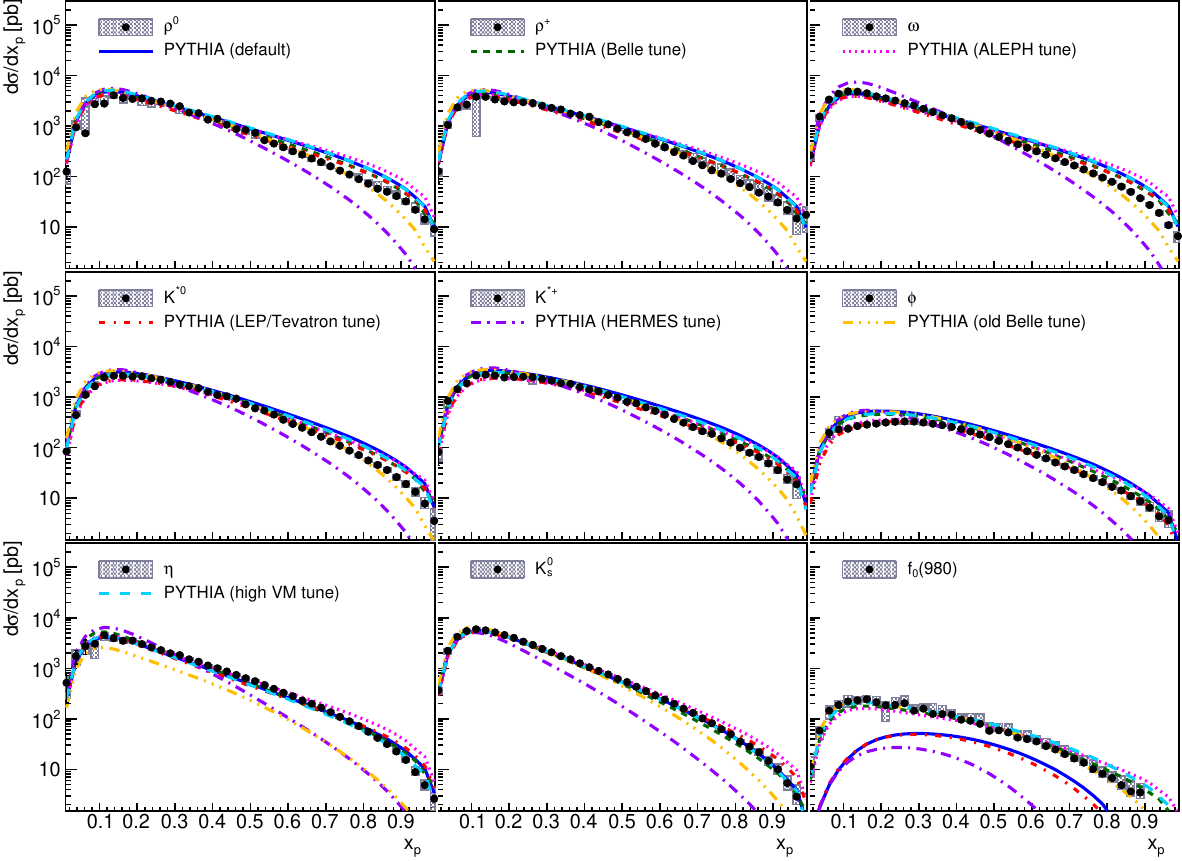}
    \caption{Production cross sections as a function of \xp for $\rho^0$, $\rho^+$, $\omega$, $K^{*0}$, $K^{*+}$, $\phi$, $\eta$, $K_S^{0}$, and $f_0(980)$ for data (black points) compared to various MC tunes as discussed in the text. }
    \label{fig:vmtunes}
\end{figure*}

\subsection{Tune comparisons}
As discussed in the previous sections about acceptance and ISR corrections, various {\sc pythia} tunes are used to estimate the dependence of these corrections on the description of the fragmentation itself. It follows that these tunes can also be compared to the final results to learn which parameters in {\sc pythia}6 do best in describing fragmentation. 
For the $D$ meson results, the most relevant parameters are the Lund-model parameters related to the mass (PARJ 42), the high-\xp behavior (PARJ 41), and the transverse momentum produced in the fragmentation (PARJ 21). In Fig.~\ref{fig:dmtunes}, all tunes display the typical peaking behavior of the inclusive cross sections, but the peak positions vary significantly. 
The {\sc hermes} tune has the highest suppression at high \xp (PARJ 41 of 1.94) and therefore peaks at the lowest values, far away from the actual data. All other tunes are not only closer together (with PARJ 41 values ranging from 0.3 to 0.5), but are generally closer to the data. Apart from the absolute magnitude, the default {\sc pythia} and Belle-related tunes reproduce the shape and peak position best, while {\sc lep} and {\sc aleph} tunes either peak at lower \xp or higher \xp, respectively. One striking exception is the behavior of the strange $D$ mesons, where only the {\sc aleph} tune is able to roughly reproduce the peak position. Furthermore, apart from the {\sc lep} tune, all tunes largely overestimate the overall magnitude. This suggests that the additional strangeness further pushes the peak to higher \xp and suppresses the production cross section \footnote{This is in contrast to strange-baryon production, where {\sc pythia} is found to underestimate hyperon production with increasing strangeness~\cite{Belle:2017caf}.}. 
Figure \ref{fig:dmratios} displays the different tunes for the charmed mesons relative to the measured cross sections.

Part of the magnitude differences between the tunes and the data for all $D$ mesons can be related to the probability of producing a charmed vector meson over a pseudoscalar meson, which is encoded in PARJ 13. From these magnitude comparisons, it appears that tunes for which this parameter ranges from PARJ 13 of 0.65 to 0.75 are preferred over the {\sc lep} tune (0.54). This is roughly consistent with the optimization efforts reported in \cite{Belle:2005mtx} where a best value of 0.59 was found. 

\begin{figure*}[htb]
    \centering
    \includegraphics[width=0.9\textwidth]{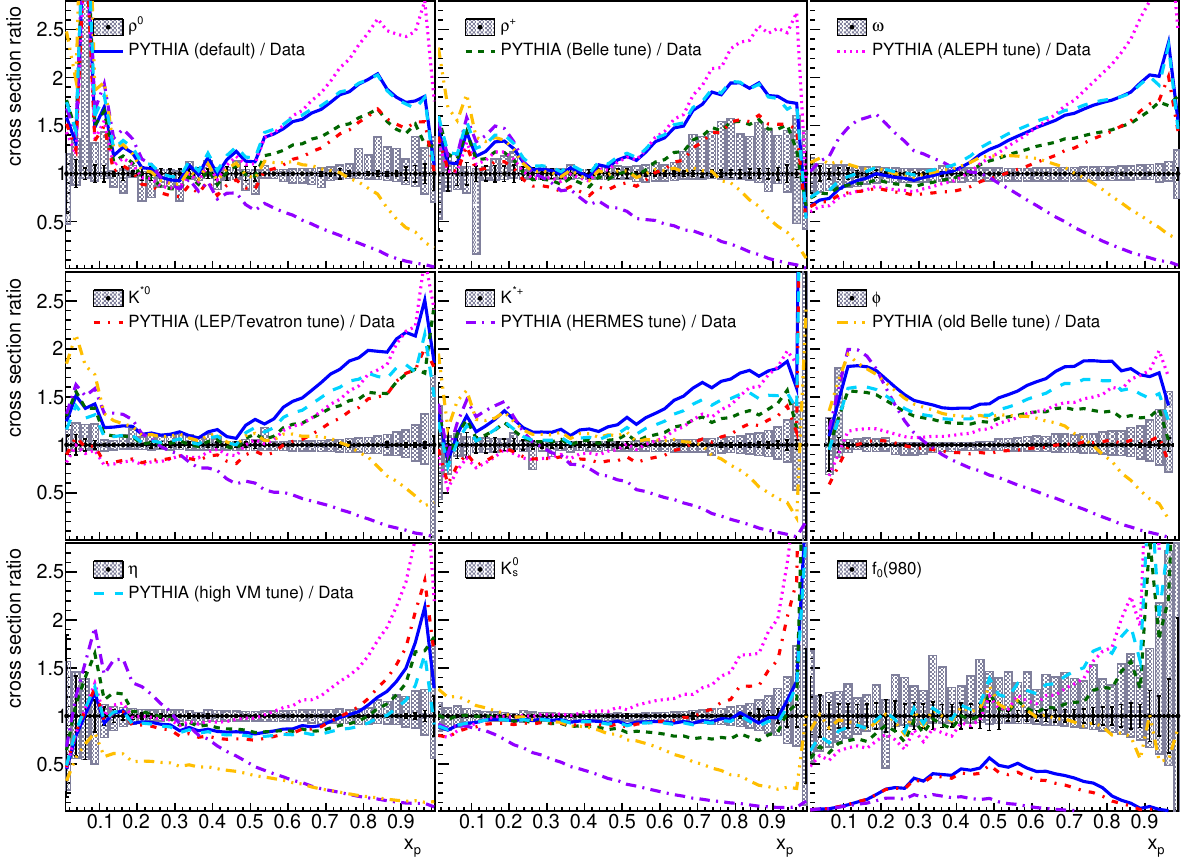}
    \caption{Production cross sections for $\rho^0$, $\rho^+$, $\omega$, $K^{*0}$, $K^{*+}$, $\phi$, $\eta$, $K_S^0$, and $f_0(980)$ from MC simulations using various fragmentation tunes (as indicated) relative to Belle data as a function of \xp. }
    \label{fig:vmratios}
\end{figure*}

Turning to the lighter mesons, more parameters are needed in the description of the fragmentation process. In the case of the vector mesons, the aforementioned ratio between vector meson and pseudoscalar meson production is again very relevant. For light quarks, this is encoded in PARJ 11 and for strange quarks in PARJ 12. The default values are 0.5 for light quarks and 0.6 for strange quarks. To study the effect of an increase in the ratio, the default {\sc pythia} setting was also tested with the light-quark ratio set to 0.6, labeled {\it high VM} in Fig.~\ref{fig:vmtunes}. Figure \ref{fig:vmratios} also displays the different tunes for the non-charmed mesons relative to the measured inclusive data. The cross sections for $\rho$ and $\omega$ mesons are indeed predicted to be slightly larger than in the default setting. However, from the measured data there does not appear to be a need to increase this fraction, which is consistent with the feed-down subtracted comparisons discussed above. 
Overall, the magnitudes, and partially the shapes, of the cross sections are described reasonably well, with the {\sc aleph}, default {\sc pythia}, and {\sc lep} tunes producing slightly harder, and the old Belle and (in particular) {\sc hermes} tunes producing softer distributions.

\section{Summary\label{sec:summary}}
In summary, we have reported the differential production cross sections in $e^+e^-$ annihilation near $\sqrt{s} = $ 10.58~GeV as a function of the momentum fraction \xp for $\rho^+$, $\rho^0$, $\omega$, $K^{*+}$, $K^{*0}$, $\phi$, $\eta$, $K_S^0$, $f_0(980)$, $D^+$, $D^0$, $D_s^+$, $D^{*+}$, $D^{*0}$, and $D^{*+}_s$. The charmed-meson results provide an update of the previous Belle measurement, which they supersede, while the $D_s^{*+}$ meson cross sections are measured for the first time in Belle.
Cross sections for vector mesons as well as those for $\eta$, and $K_S^0$ are obtained for the first time at \(B\) factories and provide additional insight into the modeling of fragmentation, particularly concerning the role of the spin of the final-state particle and its mass and fractional-momentum dependence. From these results it is concluded that no additional enhancement of spin-one fragmentation over spin-zero particles is needed in the {\sc pythia} MC event generators.
These results will also serve as valuable input for the modeling of ultra-high-energetic cosmic air showers as it relates to the impact of fragmentation on the number of detected muons at the earth's surface. 

The results will impact the use of the mesons studied here in, e.g., the investigation of spin-dependent fragmentation in semi-inclusive deep inelastic scattering, where new insights can also be gained about the spin structure of the nucleon.


\begin{acknowledgments}  

This work, based on data collected using the Belle detector, which was
operated until June 2010, was supported by 
the Ministry of Education, Culture, Sports, Science, and
Technology (MEXT) of Japan, the Japan Society for the 
Promotion of Science (JSPS), and the Tau-Lepton Physics 
Research Center of Nagoya University; 
the Australian Research Council including grants
DP210101900, 
DP210102831, 
DE220100462, 
LE210100098, 
LE230100085; 
Austrian Federal Ministry of Education, Science and Research (FWF) and
FWF Austrian Science Fund No.~P~31361-N36;
National Key R\&D Program of China under Contract No.~2022YFA1601903,
National Natural Science Foundation of China and research grants
No.~11575017,
No.~11761141009, 
No.~11705209, 
No.~11975076, 
No.~12135005, 
No.~12150004, 
No.~12161141008, 
and
No.~12175041, 
and Shandong Provincial Natural Science Foundation Project ZR2022JQ02;
the Czech Science Foundation Grant No. 22-18469S;
Horizon 2020 ERC Advanced Grant No. 884719, ERC Starting Grant No. 947006 "InterLeptons", and Grant No. 824093 "STRONG-2020" (European Union);
the Carl Zeiss Foundation, the Deutsche Forschungsgemeinschaft, the
Excellence Cluster Universe, and the VolkswagenStiftung;
the Department of Atomic Energy (Project Identification No. RTI 4002), the Department of Science and Technology of India,
and the UPES (India) SEED finding programs Nos. UPES/R\&D-SEED-INFRA/17052023/01 and UPES/R\&D-SOE/20062022/06; 
the Istituto Nazionale di Fisica Nucleare of Italy; 
National Research Foundation (NRF) of Korea Grant
Nos.~2016R1\-D1A1B\-02012900, 2018R1\-A2B\-3003643,
2018R1\-A6A1A\-06024970, RS\-2022\-00197659,
2019R1\-I1A3A\-01058933, 2021R1\-A6A1A\-03043957,
2021R1\-F1A\-1060423, 2021R1\-F1A\-1064008, 2022R1\-A2C\-1003993;
Radiation Science Research Institute, Foreign Large-size Research Facility Application Supporting project, the Global Science Experimental Data Hub Center of the Korea Institute of Science and Technology Information and KREONET/GLORIAD;
the Polish Ministry of Science and Higher Education and 
the National Science Center;
the Ministry of Science and Higher Education of the Russian Federation
and the HSE University Basic Research Program, Moscow; 
University of Tabuk research grants
S-1440-0321, S-0256-1438, and S-0280-1439 (Saudi Arabia);
the Slovenian Research Agency Grant Nos. J1-9124 and P1-0135;
Ikerbasque, Basque Foundation for Science, and the State Agency for Research
of the Spanish Ministry of Science and Innovation through Grant No. PID2022-136510NB-C33 (Spain);
the Swiss National Science Foundation; 
the Ministry of Education and the National Science and Technology Council of Taiwan;
and the United States Department of Energy and the National Science Foundation.
These acknowledgements are not to be interpreted as an endorsement of any
statement made by any of our institutes, funding agencies, governments, or
their representatives.
We thank the KEKB group for the excellent operation of the
accelerator; the KEK cryogenics group for the efficient
operation of the solenoid; and the KEK computer group and the Pacific Northwest National
Laboratory (PNNL) Environmental Molecular Sciences Laboratory (EMSL)
computing group for strong computing support; and the National
Institute of Informatics, and Science Information NETwork 6 (SINET6) for
valuable network support.

\end{acknowledgments}  


\bibliography{main}   

\appendix

\begin{table*}
\begin{center}
\caption{Parameters for the various tunes used in the comparison to the measured cross sections reported here. Each row corresponds to a {\sc Pythia/JetSet} setting that differs for at least one of the tunes. Empty entries indicate the default setting. The tune labeled "High VM" has an increased fraction of light quarks fragmenting into vector over pseudoscalar mesons compared to the default setting. A brief explanation of each parameter is also given, for details see Ref.~\cite{Sjostrand:2001yu}. The central values for the acceptance and ISR corrections were based on the Belle tune.
\label{tab:jetsettable}}
\scalebox{1.0}{
\begin{tabular}{rc c c c c c c c}
& {\sc pythia} default & Belle & {\sc aleph} & {\sc lep}/Tevatron & {\sc hermes}  & old Belle & High VM & explanation\\ \hline
PARJ(1) &0.1 & & 0.106& 0.073& 0.029& &  & diquark suppression  \\ \hline
PARJ(2) &0.3 & & 0.285& 0.2& 0.283&  && strange suppression\\ \hline
PARJ(3) &0.4 & & 0.71& 0.94& 1.2&  & & strange diquark suppression\\ \hline
PARJ(4) &0.05 & & & 0.032&   & & & spin 1 diquark suppression \\ \hline
PARJ(11) &0.5 & & 0.55& 0.31& & & 0.6 & light quark spin 1 meson probability\\ \hline
PARJ(12) &0.6 & & 0.47& 0.4& &  & & strange quark spin 1 meson probability\\ \hline
PARJ(13) &0.75 & & 0.65& 0.54& & & & heavy quark spin 1 meson probability\\ \hline
PARJ(14) & 0& & 0.02& & & 0.05 & 0.05&  S 0, L 1, J 1 \\ \hline
PARJ(15) & 0& & 0.04& & & 0.05  & 0.05&  S 1, L 1 J 0\\ \hline
PARJ(16) & 0& & 0.02& & & 0.05  & 0.05&  S 1, L 1 J 1\\ \hline
PARJ(17) & 0& & 0.2 & & & 0.05 & 0.05 &  S 1, L 1 J 2\\ \hline
PARJ(19) & 1& & 0.57& & &  & & extra baryon suppression\\ \hline
PARJ(21) &0.36 & & 0.37& 0.325& 0.4& 0.28 & & $p_x,p_y$ \\ \hline
PARJ(25) &1 & & & 0.63& & 0.27 & & $\eta$ suppression\\ \hline
PARJ(26) &0.4 & & 0.27& 0.12& & & & $\eta'$ suppression\\ \hline
PARJ(33) &0.8& & & & 0.3&  && end of fragmentation energy\\ \hline
PARJ(41) &0.3 & & 0.4& 0.5& 1.94& 0.32& & Lund parameter $a$ \\ \hline
PARJ(42) &0.58 & & 0.796& 0.6& 0.544& 0.62& & Lund parameter $b$\\ \hline
PARJ(45) &0.5 & & & & 1.05&  & & $a$ factor for diquarks\\ \hline
PARJ(47) &1. & & & 0.67& &  &  & heavy quark endpoint modification\\ \hline
PARJ(54) & -0.050&-0.04 & -0.04& & & & &charm FF parameterization \\ \hline
PARJ(55) & -0.005&-0.004 & -0.0035& & & &&bottom FF parameterization \\ \hline
PARJ(81) &0.29 & & 0.292& 0.29& & & &$\Lambda$ value in running $\alpha_s$ \\ \hline
PARJ(82) &1.0 & & 1.57& 1.65 & & &&mass cut-off for parton showers\\ \hline
MSTJ(11) &4 & & 3& 5& & 4  & & type of FF model\\ \hline
MSTJ(12) &2 & & 3& & 1&  &&baryon production model\\ \hline
MSTJ(26) & 2& 0 & & & & 0 & &$B$-$\bar{B}$ mixing\\ \hline
MSTJ(45) & 5 & & & & 4&   & &max. flavor in gluon shower\\ \hline
MSTJ(107) & 0&1 & & & &1  & & radiative corrections\\ \hline
\end{tabular}

}
\end{center}
\end{table*}

\begin{table*}[htb]
    \centering
    \caption{Branching fractions for the various decay modes of particles analyzed and reported here \cite{ParticleDataGroup:2024cfk}. The branching fraction of the $f_0(980)$ is currently unknown and is assumed to be 52\% as implemented in the Belle MC. The Clebsch-Gordan coefficients for the $K^*$ decays into charged kaons are treated in the acceptance correction, as mentioned in the text. \label{tab:BRs}
}
    \begin{tabular}{r  r  r}
  Particle &   Particle decay &  \multicolumn{1}{r}{Branching fractions} \\ \hline 
$K_S^0 $&$ \pi^+\pi^-$   & $0.6920\pm0.0005$  \\ 
$\rho^0 $&$ \pi^+\pi^-$  & $1.0\pm0$ \\
$\rho^\pm $&$ \pi^\pm\pi^0 \rightarrow\pi^\pm (\gamma\gamma)$       & $(1.0)$ $\times$ 0.98823 \\
$f_0(980)$&$ \pi^+\pi^-$        & 0.52 \\
$K^{*0} $&$ \pi K$          & $0.99754\pm0.00021$ \\
$K^{*+} $&$ \pi K$        & $0.99902\pm0.00009$ \\
$D^0$&$ \pi^+ K^-$              & $0.03947\pm0.0003$ \\
$\phi $&$ K^+K^- $              & $0.491\pm0.005$ \\\hline
$\eta $&$ \pi^+\pi^-\pi^0\rightarrow\pi^+\pi^- (\gamma\gamma)$      & $(0.2302\pm0.0025)$ $\times$ 0.98823  \\
$\omega $&$ \pi^+\pi^-\pi^0\rightarrow\pi^+\pi^- (\gamma\gamma)$    & $(0.892\pm0.007)$ $\times$ 0.98823  \\
$D^+$&$ K^-\pi^+\pi^+$          & $0.0938\pm0.0016$  \\ 
$D_s^{+}$&$ \phi \pi^+ \rightarrow (K^+K^-)\pi^+$         & $0.0221\pm0.0006$ \\ 
$D_s^{*+}$&$ \phi \pi^+ \gamma\rightarrow (K^+K^-)\pi^+ \gamma$         & $(0.936\pm0.007)$ $\times$ $(0.0221\pm0.0006)$ \\ 
\hline
$D^{*+}$&$ D^0 \pi^+ \rightarrow (K^-\pi^+)\pi^+$           & $(0.677\pm0.005)$ $\times$ 0.03947  \\
$D^{*0}$&$ D^0 \pi^0 \rightarrow (K^-\pi^+)(\gamma\gamma)$           & $(0.647\pm0.009)$ $\times$ 0.03947$\times$ 0.98823 \\ \hline
\end{tabular}
\end{table*}

\end{document}